%% file: neurips_2026.tex
\documentclass{article}

  \usepackage[main, final]{neurips_2026}

\usepackage[utf8]{inputenc} 
\usepackage[T1]{fontenc}    
\usepackage{hyperref}       
\usepackage{url}            
\usepackage{booktabs}       
\usepackage{amsfonts}       
\usepackage{nicefrac}       
\usepackage{microtype}      
\usepackage{xcolor}         

\usepackage{xspace}
\usepackage{hyperref}
\usepackage{xspace}
\usepackage{tcolorbox}
\tcbuselibrary{skins}
\usepackage[table]{xcolor}
\usepackage{multirow}
\usepackage{wrapfig}
\usepackage[normalem]{ulem}

\usepackage{amsmath}
\usepackage{amssymb}
\usepackage{mathtools}
\usepackage{amsthm}
\newcommand{\SMark}{\textsc{SumMark}\xspace}

\title{Auditor-Assisted Summary-Channel Verification for Hosted LLM Identity Substitution}

\author{%
  Ziyi Zhang\\
  Stony Brook University\\
  \texttt{ziyi.zhang.3@stonybrook.edu} \\
  \And
  Ziyao Wang \\
  University of Maryland, College Park\\
  \texttt{ziyaow@umd.edu } \\
   \And
  Guoheng Sun \\
  University of Maryland, College Park \\
   \texttt{ghsun@umd.edu} \\
   \And
  Ang Li \\
  University of Maryland, College Park \\
   \texttt{angliece@umd.edu} \\
   \And
  Jian Li\\
  Stony Brook University \\
   \texttt{jian.li.3@stonybrook.edu} \\
}

\begin{document}

\maketitle

\begin{abstract}
Hosted LLM services increasingly expose model identity as a product claim, but external users and auditors often cannot inspect which model configuration served a request. We study an auditor-assisted regime in which a trusted auditor enrolls claimed identities and prepares verification material before deployment, while audit-time verification uses only the claimed identity and released output, without runtime access to model internals, inference logs, routing decisions, or provider metadata. We propose \SMark, a \emph{summary-channel watermarking} framework that uses user-visible reasoning summaries as a keyed audit surface. During setup, \SMark constructs identity-specific vocabulary carriers, trains lightweight summary adapters, and calibrates per-identity thresholds. At deployment time, a keyed statistical test assesses whether the released summary is consistent with the claimed identity and, when needed, attributes the most likely enrolled identity. Across model families, scales, and task domains, \SMark reliably detects cross-family and cross-scale substitutions at low false-positive rates while preserving task-facing answer quality. We further study FP16$\rightarrow$INT8/INT4 precision-consistency diagnostics and robustness under bounded post-processing, rewriting stress tests, and text-only spoofing, revealing both strong performance and clear failure boundaries. Our code is publicly available at \url{https://github.com/trinitr0toluene/Summark}.
\end{abstract}

\input{intro2}

\input{prelim2}

\input{method2}

\input{exp2}

\section{Conclusion}

We introduce \SMark, a summary-channel watermarking framework for auditing hosted LLM identity claims from user-visible reasoning summaries. \SMark embeds keyed, identity-specific signals into summaries, enabling low-FPR substitution detection and multi-class attribution while preserving task-facing answer quality. Our results also characterize important boundaries: precision-consistency diagnostics are useful but model-dependent, bounded post-processing is well tolerated, and strong rewriting or adaptive spoofing remains challenging. Future work includes richer audit signals beyond text and stronger robustness for fine-grained deployment variants.

\input{ack}


\bibliographystyle{plain}
\bibliography{example_paper}






\clearpage
\newpage
\appendix



\input{appendix}



%

\end{document}

%% file: intro2.tex
\section{Introduction}\label{sec:intro}

Hosted Large Language Model (LLM) services increasingly offer \emph{model selection} as a product feature: users or enterprise customers request a named model family and parameter scale, and receive responses through a standardized interface. This creates an incentive for cost-motivated \emph{identity substitution}, where a provider continues to claim the requested model identity while serving responses through a different and often cheaper model configuration. Detecting such behavior from final answers alone is difficult because related models can produce superficially similar outputs for many prompts, and because surface style can be smoothed by prompting, decoding, or response normalization~\cite{cai2025you}. This motivates externally checkable evidence for auditing claimed identities in hosted LLM services, especially when deployment-time access to provider internals is unavailable.

\begin{figure}[t]
    \centering
    \includegraphics[width=0.95\linewidth]{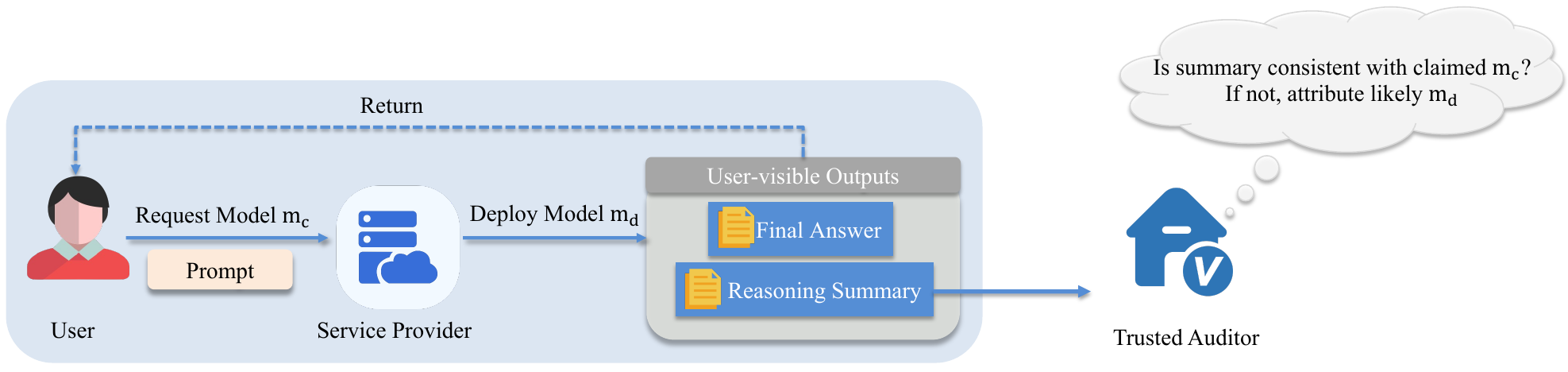}
  \caption{\textbf{Auditor-assisted summary-channel verification.} The user requests a claimed model identity $m$ and receives a task-facing answer with a user-visible reasoning summary. Given the claim and released summary, a trusted auditor tests whether the summary is statistically consistent with $m$ using secret verification material prepared offline. When answer and summary generation use separate pipelines, the guarantee is summary-channel consistency. End-to-end answer provenance requires additional coupling or deployment-side attestation.}
    \label{fig:intro}
\end{figure}

Existing approaches occupy two broad extremes. At one end, pure output-behavior audits require no privileged setup or runtime access: they query the service and try to infer identity from response behavior, stylistic fingerprints, or challenge-response signatures~\cite{cai2025you,gao2024model,pasquini2025llmmap,he2018verideep}. However, such evidence is unkeyed, prompt-dependent, and brittle under response variability, adaptive routing, provider-side normalization, or paraphrasing~\cite{cai2025you,krishna2023paraphrasing}. At the other end, infrastructure-level verification can provide stronger evidence by relying on trusted instrumentation, execution traces, inference logs, cryptographic or trusted-execution mechanisms, or deployment-side attestation~\cite{sun2024svip,tramer2018slalom,chen2024zkml}. These mechanisms are powerful when available, but they require per-request runtime visibility into the provider-controlled serving stack, which may be unavailable to external auditors and can conflict with privacy, safety, or proprietary constraints~\cite{casper2024black}.

We study the middle ground between these extremes, which we call \emph{onboarded-but-runtime-blind auditing}. In this regime, a trusted auditor enrolls a finite set of claimed identities and prepares secret verification material during a one-time offline setup phase, but deployment-time auditing must operate from public response fields without runtime access to model internals, inference logs, routing decisions, or provider-reported metadata. Offline enrollment defines the configured identities; runtime auditing checks whether a returned auditable field is statistically consistent with the claimed identity. This regime captures settings such as third-party audits, regulatory evaluations, contractual reviews, or provider-volunteered transparency programs, where certified identities may be exposed during setup while the production serving stack remains unavailable on every request.

In this paper, we focus on one user-visible field as the audit carrier: \emph{reasoning summaries} accompanying final answers. Reasoning summaries are increasingly exposed by hosted reasoning services as concise natural-language explanations while full chains-of-thought remain hidden. For example, OpenAI's reasoning models~\cite{openai2024} and Google's Gemini thinking models~\cite{gemini2024} support summarized reasoning. These summaries are observable at deployment time, less tied to task correctness than final answers, and separate from the task-facing answer the user relies on. Our choice of summaries is therefore a carrier-design choice: summaries provide a user-visible, semantically related, and flexible field where audit evidence can be placed without directly perturbing the task-facing answer.

A natural way to turn this field into externally checkable evidence is through \emph{watermarking}, applied to reasoning summaries rather than task-facing answers. Existing LLM watermarks typically embed key-dependent statistical biases into generated text and detect them through token-frequency or likelihood-based tests~\cite{kirchenbauer2023watermark,zhao2023provable,dathathri2024scalable,liu2024survey}. When applied to final answers, however, the watermark carrier is tightly coupled to prompt semantics, formatting constraints, lexical entropy, and downstream post-processing. This can dilute detectability, create utility tradeoffs, and increase sensitivity to paraphrasing or rewriting~\cite{piet2025markmywords,rastogi2024revisiting,jovanovic2024watermark,chen2024mark}. In contrast, \emph{summary-channel watermarking} places keyed audit evidence in a separate user-visible field, enabling deployment-time verification while leaving task-facing answers minimally affected. This perspective leads to the following question:
\vspace{-0.1in}
\begin{tcolorbox}[colback=white!5!white,colframe=white!75!white]
\textit{Can the released reasoning summary serve as a keyed measurement surface for deployment-time identity-consistency checks in an onboarded-but-runtime-blind hosted LLM service, without watermarking task-facing answers or requiring runtime access to internal traces?}
\end{tcolorbox}
\vspace{-0.1in}
We introduce \SMark, a \emph{summary-channel watermarking} framework for auditor-assisted identity-consistency verification over a finite, pre-enrolled set of identities (Figure~\ref{fig:intro}). During offline setup, the auditor uses a secret key to construct identity-specific vocabulary buckets, train lightweight summary watermark adapters, and calibrate per-identity decision thresholds. At deployment time, given a claimed identity and a released reasoning summary, the auditor applies a keyed statistical test to assess identity consistency. The same statistic supports both detection of identity-changing substitutions and attribution of the most likely generating identity among enrolled candidates. The deployment-time verifier requires no access to model parameters, activations, hidden traces, inference logs, runtime routing decisions, or provider-reported metadata.

This formulation defines the verification target of \SMark. The verifier provides statistical evidence that the released reasoning-summary field is consistent with the claimed enrolled identity under the auditor-assisted setup above. \SMark is therefore a deployment-time summary-channel audit layer: stronger than unkeyed behavioral testing over ordinary outputs, but complementary to infrastructure-level attestation when full serving provenance is required.
Our evaluation shows that this summary-channel signal supports reliable low-FPR detection and attribution for identity-changing substitutions across model family and scale, while preserving task-facing answer quality. We further use post-processing, precision-variant, and spoofing stress tests to characterize where the signal remains stable and where it begins to degrade.
Our main contributions are as follows:

$\bullet$ \textbf{Auditor-assisted verification regime.} 
We formulate an onboarded-but-runtime-blind regime for hosted LLM identity claims: a trusted auditor enrolls identities and prepares keyed verification material offline, while audit-time decisions use released summaries without runtime access to provider logs, traces, routing decisions, model internals, or self-reported metadata.

$\bullet$ \textbf{Reasoning summaries as a keyed audit surface.} 
We identify the user-visible reasoning-summary field as a practical audit carrier that is externally observable at deployment time and can hold verification evidence while isolating the task-facing answer from watermark perturbation.

$\bullet$ \textbf{Identity-specific summary-channel watermarking.} 
We instantiate this idea in \SMark, which uses keyed, identity-specific vocabulary carriers and lightweight summary adapters to support low-FPR substitution detection and multi-class attribution over enrolled identities.

$\bullet$ \textbf{Boundary-aware evaluation.} 
Across model families, scales, and task domains, we evaluate \SMark under identity-changing substitutions, attribution, answer-quality preservation, behavioral and generic watermarking baselines, precision-variant diagnostics, and robustness stress tests. The results show strong low-FPR summary-channel verification while exposing clear boundaries under stronger post-processing and spoofing.

%% file: prelim2.tex
\section{Threat Model}\label{sec:prelim}

We consider an auditor-assisted model identity auditing setting with three roles. A \emph{service provider} hosts an LLM service and exposes a user-facing interface; an \emph{end user} issues prompts and consumes responses; and a \emph{trusted auditor} verifies model identity on behalf of users, customers, or regulators. For each request, the provider returns (i) a task-facing final answer and (ii) a user-visible reasoning summary, while presenting the response under a claimed identity $m_c$. We define a model identity by its $(\text{family},\text{parameter scale})$ pair. Numerical precision is treated separately as a within-identity deployment variant rather than as a base identity. The end user holds no secret material and interacts solely through the public interface. At deployment time, the auditor observes only the claimed identity $m_c$ and the released reasoning summary $s$.

\textbf{Auditor access and certification target.}
The auditor's access is split into two phases. During a one-time \emph{offline setup}, the auditor has trusted reference access to enrolled identities, sufficient to prepare per-identity verification material and calibration thresholds; concrete procedures are described in Section~\ref{sec:method}. During \emph{deployment-time auditing}, the auditor's decision depends only on $(m_c,s)$ and a secret verification key. The auditor does not observe model parameters, internal activations, hidden reasoning states, inference logs, runtime routing decisions, or provider-reported metadata.

The audit target is the released reasoning-summary channel. Given $(m_c,s)$, the auditor tests whether $s$ is statistically consistent with the claimed identity under the auditor's keyed verification material. This summary-only decision does not, by itself, certify the provenance of the task-facing answer when answer generation and summary generation are decoupled. End-to-end serving provenance requires additional answer-summary coupling, deployment policy, or infrastructure-level attestation, which is complementary to the statistical audit layer studied here.

\textbf{Provisioning and trust assumptions.}
The auditor's verification material is provided to the serving stack only through authenticated invocation: the provider may use the authorized material during legitimate serving, but cannot extract the secret key, carrier sets, adapter weights, or reusable offline material for other identities. This invocation channel may be instantiated by an auditor-hosted service, a protected serving wrapper, or another authenticated provisioning mechanism. \SMark evaluates the statistical audit layer under this access model rather than proposing a full enforcement protocol. This setting is intended for provider-participating audits, such as external audits, procurement checks, regulatory reviews, or voluntary transparency programs.
It is stronger than pure API-only access but weaker than full runtime attestation. We assume the auditor's key store is not compromised and the provisioning channel is authenticated.

\textbf{Adversary (provider).}
The provider may be cost-motivated and may attempt to make the auditor accept the claimed identity $m_c$ even when the auditable summary channel is generated under, or made to appear consistent with, a different served identity $m_d\neq m_c$. A stronger form of evasion is \emph{spoofing}, in which the provider tries to reproduce the verification signal expected under $m_c$ using a substituted model. We consider the following capabilities.

\emph{Identity substitution.}
Let $\mathcal{M}$ be a finite set of enrolled identities at the family$\times$scale granularity. The adversary may perform \emph{cross-scale substitution}, where the served identity changes parameter scale within the same model family; \emph{cross-family substitution}, where the served identity changes model family; or \emph{joint family-and-scale substitution}, where both attributes differ, e.g., serving Qwen-1.5B while claiming LLaMA-3B. These are the primary identity-substitution events studied in this paper.

\emph{Precision variants.}
Our primary identity universe treats model identity as a family-scale pair. Numerical precision changes, such as FP16$\rightarrow$INT8 or FP16$\rightarrow$INT4, preserve this identity and are therefore not counted as identity-substitution events. We evaluate them separately as \emph{precision-consistency diagnostics}: for a fixed identity $m$, the verifier calibrated on FP16 summaries tests whether summaries from a quantized deployment remain consistent with that calibration. Mechanism details and results appear in Sections~\ref{sec:method} and~\ref{sec:exp:precision}.

\emph{Summary post-processing.}
The primary post-processing model allows \emph{bounded online transformations} to the released summary, such as formatting edits, deletion, truncation, or sentence reordering. These transformations preserve the released text as the reasoning-summary field for the same request and test whether the verifier depends on exact phrasing or sentence order. Stronger transformations, such as full LLM-based paraphrase rewriting or fine-tuning-based watermark removal, are not part of the primary bounded post-processing model because they regenerate or substantially alter the summary rather than preserving it up to surface edits. We evaluate them as stress tests in Section~\ref{sec:exp:robustness} to characterize the boundary of text-based summary verification.

\emph{Watermark spoofing.}
The provider may observe or query watermarked summaries produced under the claimed identity, but does not obtain the auditor's secret key, carrier sets, adapter weights, or reusable offline access to the claimed identity's verification material. The adversary may then try to reproduce $m_c$'s verification signal on a substituted model $m_d$ from text-only observations. We evaluate two spoofing strategies in Section~\ref{sec:exp:robustness}: \emph{adapter distillation}, which trains a local generator on $m_d$ from observed prompt-summary pairs, and \emph{watermark stealing}, which estimates token-level signal statistics from watermarked texts and biases decoding accordingly. These attacks test whether the claimed verification signal can be forged from text-only observations alone.

%% file: method2.tex
\section{\SMark Framework}\label{sec:method}

\begin{figure*}[t]
    \centering
    \includegraphics[width=0.95\linewidth]{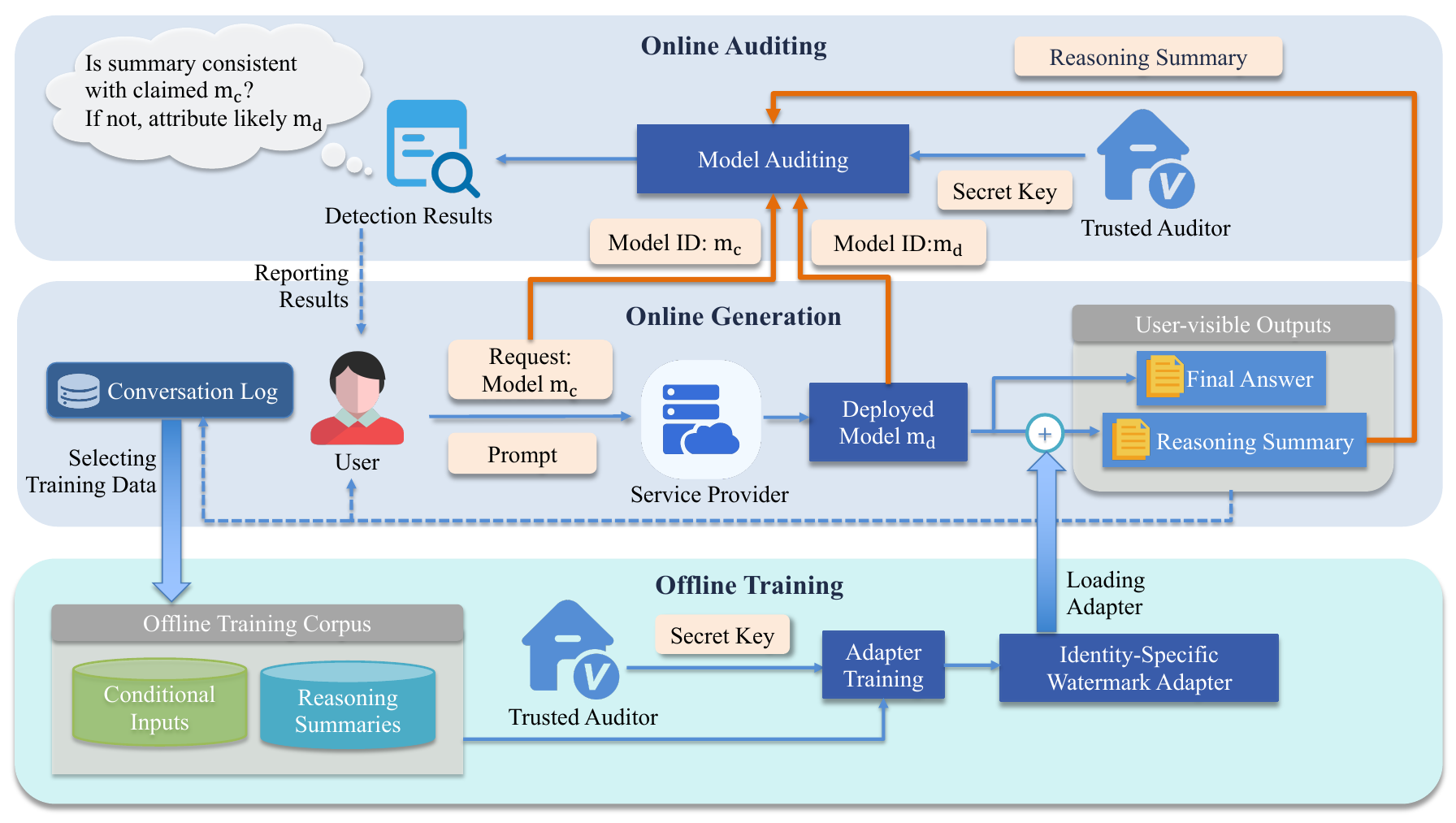}
    \caption{\textbf{Workflow of \SMark.} Offline, the trusted auditor uses a secret key to construct identity-specific vocabulary carriers, train summary watermark adapters, and calibrate thresholds. Online, the provider returns a task-facing answer and reasoning summary under a claimed identity. The auditor uses only the claim, released summary, and secret key to test summary-channel consistency and, when needed, attribute the most likely enrolled identity.}
    \label{fig:framework}
\end{figure*}

We present \SMark, a summary-channel framework for auditor-assisted identity-consistency verification (Figure~\ref{fig:framework}). The central idea is to \textit{turn the released reasoning summary into a keyed measurement surface.} Each enrolled identity receives a secret vocabulary carrier, and a lightweight summary adapter imprints that carrier into its summaries. At deployment time, the verifier receives only the claimed identity $m_c$, released summary $s$, and auditor's secret key, and tests whether the summary's token statistics match the carrier calibrated for $m_c$.

\SMark has three components. \textit{First}, the auditor constructs identity-specific carrier sets $\{G_m\}_{m\in\mathcal M}$ with controlled density and limited cross-identity overlap. \textit{Second}, for each enrolled identity, the auditor trains a lightweight adapter $\Delta_m$ that increases the corresponding carrier footprint while preserving fluency and faithfulness. \textit{Third}, the auditor calibrates thresholds and performs a keyed test for detection or attribution using only the released summary. Unlike answer-level watermarking, \SMark localizes audit evidence to the reasoning summary. The final answer is the task-facing output, so watermarking it can create utility or formatting tradeoffs. The reasoning summary is an auxiliary user-visible field with more lexical slack, allowing \SMark to embed identity-specific evidence while minimally affecting the task-facing answer.

\subsection{Problem Definition}\label{sec:probdef}

Let $\mathcal M$ be a finite set of enrolled identities, where each $m\in\mathcal M$ specifies a model family and parameter scale. During offline setup, the auditor prepares identity-specific verification material for each~$m$: a carrier set $G_m\subseteq V$ over the vocabulary $V$, a summary watermark adapter $\Delta_m$, and a detection threshold calibrated
for a target false-positive rate. At deployment time, for a user prompt $x$, the provider returns a response $r=(y,s)$ consisting of a task-facing answer $y$ and a user-visible reasoning summary $s$, while presenting the response under a claimed identity $m_c\in\mathcal M$. The verifier receives only $(m_c,s)$ and the auditor's secret key; it does not access model parameters, activations, logits, hidden reasoning states, inference traces, runtime routing decisions, system logs, or provider-reported metadata.

\textbf{Detection.}
Given $(m_c,s)$, the verifier tests whether the released summary is statistically consistent with the claimed identity $m_c$ or is more consistent with a different enrolled identity $m_d\neq m_c$.

\textbf{Attribution.}
Given $s$ and~$\mathcal M$, identity attribution selects the enrolled identity whose keyed carrier statistic best explains the released summary.

\textbf{Substitution types.}
We consider three identity-changing substitution categories over $\mathcal M$: (i) \emph{cross-family substitution}, where the served identity changes model family; (ii) \emph{cross-scale substitution}, where the served identity changes parameter scale within the same family; and (iii) \emph{joint family-and-scale substitution}, where both attributes differ, e.g., serving Qwen-1.5B while claiming LLaMA-3B.

\textbf{Precision variants.}
Our primary task treats identities as family-scale pairs. We separately study whether the FP16-calibrated verifier for a fixed identity $m$ provides useful consistency evidence when the same identity is served at INT8 or INT4 precision. This is not an attribution task over $\mathcal M$: all precision variants reuse the same carrier set $G_m$ and adapter $\Delta_m$, and the verifier flags a variant only when its summary statistics deviate from FP16 calibration.


\subsection{Identity-Specific Summary Carriers}\label{sec:wm_design}

\textbf{Core idea.}
\SMark assigns each identity $m\in\mathcal M$ a secret, key-derived subset of vocabulary tokens $G_m\subseteq V$, which we call its \emph{carrier set}. During summary generation, the provider applies the identity-specific adapter $\Delta_m$ to nudge token usage toward $G_m$ while maintaining a fluent and faithful summary. This produces an identity-dependent footprint that is detectable from summary text alone using the auditor's key.

\textbf{Audit-time statistic.}
Let $s=(s_1,\ldots,s_T)$ be the released summary. For identity $m$, the verifier computes the green-match count
$C_m(s)=\sum_{t=1}^{T}\mathbb{I}[s_t\in G_m],$
and converts it to a calibrated score $Z_m(s)=(C_m(s)-\mu_m(T))/\sigma_m(T)$ using held-out summaries for identity $m$. Detection compares $Z_{m_c}(s)$ to a per-identity threshold chosen for a target false-positive rate, while attribution selects the identity with the largest score. Full calibration details are in Appendix~\ref{sec:detection}.

\textbf{Constructing many separable carrier sets.}
A single global greenlist would create a shared signal across all models and does not support reliable attribution. \SMark instead uses the secret key $K$ to construct a family of carrier sets $\{G_m\}_{m\in\mathcal M}$ with two tunable properties:

$\triangleright$ \textbf{Controlled green ratio.}
Each identity receives a carrier set of comparable density:
\begin{equation}\label{eq:green_ratio}
    {|G_m|}/{|V|}\approx \rho,\quad \forall m\in\mathcal M,
\end{equation}
which stabilizes the null baseline across identities.

$\triangleright$  \textbf{Controlled cross-identity overlap.}
The construction also limits cross-identity collisions:
\begin{equation}\label{eq:overlap}
    \frac{|G_m\cap G_{m'}|}{|G_m|}\le \tau,\quad \forall m\neq m',
\end{equation}
so different identities remain distinguishable even when the auditor sees only short summaries.

The pair $(\rho,\tau)$ governs the trade-off between watermark strength and identity capacity: larger $\rho$ increases per-token evidence, while smaller $\tau$ improves separability across identities. This differs from a shared watermark: a shared greenlist can indicate that some watermark is present, but cannot distinguish which enrolled identity produced the summary. Controlled overlap is what turns watermark detection into model-identity attribution. We provide a concrete keyed construction and a capacity characterization in Appendix~\ref{app:bucket}.


\subsection{Training Summary Watermark Adapters}\label{sec:lora}

\textbf{Embedding mechanism.}
To realize an identity-dependent bias in the summary channel, \SMark trains a lightweight adapter for each enrolled identity. For each identity $m\in\mathcal M$ with base weights $\theta_m$, we train a summary adapter $\Delta_m$ while keeping $\theta_m$ frozen. At deployment time, the provider enables $\Delta_{m_c}$ only when generating the reasoning summary for a request that claims identity $m_c$.

\textbf{Training objective.}
Adapter training balances three goals:
(i) \emph{utility}, so summaries remain fluent and faithful;
(ii) \emph{detectability}, so the summary exhibits an elevated footprint of $G_m$; and
(iii) \emph{separability}, so the correct identity's footprint exceeds competing identities' footprints.
We instantiate the base identity objective as
\begin{equation}\label{eq:loss_id}
    L=L_{\text{sft}}+\lambda_{\text{wm}} L_{\text{wm}}+\lambda_{\text{sep}} L_{\text{sep}}.
\end{equation}
Here, $L_{\text{sft}}$ is a standard supervised fine-tuning loss on reference summaries, $L_{\text{wm}}$ encourages increased usage or probability mass of the identity-specific carrier set $G_m$, and~$L_{\text{sep}}$ penalizes confusion by enforcing a margin between the footprint of $G_m$ and competing carrier sets. Intuitively, $L_{\text{wm}}$ makes the claimed identity detectable, while $L_{\text{sep}}$ makes it distinguishable from other enrolled identities. Full definitions are provided in Appendix~\ref{app:training}.

\textbf{Optional precision-consistency regularizer.}
For the sidecar precision analysis described in Section~\ref{sec:probdef}, \SMark reuses the same carrier set $G_m$ and the same adapter $\Delta_m$ across the FP16, INT8, and INT4 variants of identity $m$. We do not introduce precision-specific carrier sets or precision-specific adapters. Instead, to make the FP16-calibrated verifier sensitive to precision-induced changes, we optionally add a lightweight regularizer that increases the separation between the carrier statistic observed under FP16 forward passes and the statistic observed under simulated quantized forward passes of the same identity. This regularizer does not create separate precision identities; it only makes the single identity-level verifier more informative for precision-consistency diagnostics. The full objective is provided in Appendix~\ref{app:training}, {and Appendix~\ref{app:ablations}} ablates its effect.

\textbf{Why adapters?}
Adapters make the watermark (a) cheap to provision and switch per identity, (b) tied to the enrolled model family and scale rather than a generic decoding-time bias, and (c) localized to summary generation so task-facing answers are minimally affected. LoRA adapters are shape-bound to enrolled bases, making cross-family or cross-scale cross-loading incompatible in our setting; same-shape precision variants remain compatible and are treated as diagnostics. Our framework is not tied to LoRA specifically: any mechanism that induces controlled, identity-dependent token preferences in summary generation is compatible.

\begin{tcolorbox}[
    enhanced, 
    width=\textwidth,
    arc=3mm, 
    boxrule=1pt, 
    colback=black!5, 
    colframe=black!75!black, 
    fonttitle=\bfseries, 
]
\noindent{\textbf{Takeaway.}
Taken together, the carrier construction and adapter objective turn ordinary watermark detection into identity-specific verification. A generic watermark can indicate that a text was watermarked, but not 
which enrolled identity it is consistent with. \SMark's key contribution is to make the summary watermark separable across identities and auditable from text alone at strict false-positive rates.}
\end{tcolorbox}

%% file: exp2.tex
\section{Experiments}\label{sec:exp}

We organize the evaluation around five claims: reliable low-FPR substitution detection, multi-class attribution, superiority over behavioral and generic watermarking baselines, task-facing answer-quality preservation, and boundary behavior under precision variants, post-processing, rewriting, and spoofing. Section~\ref{sec:mainresults} addresses detection, attribution, and answer quality; Section~\ref{sec:baselines} isolates the role of identity-specific summary watermarking; and Sections~\ref{sec:exp:precision}--\ref{sec:exp:robustness} study boundary cases. {Appendix~\ref{app:ablations}} provides ablations on watermark design and training loss components.
We further include a qualitative case study in Appendix~\ref{sec:case_study}, showing how the adapter leaves an identity-specific token pattern in the summary text.

\subsection{Experimental Settings}\label{sec:exp_setup}

\textbf{Models and datasets.}
We evaluate four enrolled identities: Qwen2.5-1.5B, Qwen2.5-3B, Qwen2.5-7B, and LLaMA3.2-3B. Appendix~\ref{app:full_detection} extends this to 12 identities across four model families, with a separate budget of 1,000 examples per identity. We use three datasets that induce distinct summary styles: Math for mathematical reasoning, Code for programming, and General for general problem solving. These domains differ in both task content and summary structure: Code summaries tend to be more templated and procedural, while General summaries are more stylistically varied and conversational. Full dataset details appear in Appendix~\ref{app:datasets}.

\textbf{Substitution and precision-variant settings.}
An identity~$m\in\mathcal M$ is a family-scale pair. In the primary substitution benchmark, the provider serves a different identity while claiming~$m_c$, covering cross-scale, cross-family, and joint family-and-scale substitutions. Unless stated otherwise, the substitution ratio is 50\%. 
Separately, we evaluate \emph{precision-consistency diagnostics}: the claimed configuration is~$(m,\mathrm{FP16})$, while the served configuration is~$(m,\mathrm{INT8})$ or~$(m,\mathrm{INT4})$ for the same identity~$m$. Precision variants reuse the same carrier set~$G_m$ and adapter~$\Delta_m$, so they are evaluated as within-identity consistency checks rather than multi-class attribution over~$\mathcal M$.

\textbf{Baselines.}
We compare against two classes of baselines. First, \emph{text-classifier baselines} operate on released summaries without watermarking: BERT~\cite{devlin2019bert}, GPT-2~\cite{radford2019language}, and T5~\cite{raffel2020exploring} are trained to predict consistency with a claimed identity from summary text alone. These baselines test whether unkeyed surface or stylistic cues are sufficient for identity auditing. Second, \emph{watermark baselines} operate on the same summary channel: Plain (no watermark), Greenlist (one shared greenlist across identities), Unigram~\cite{zhao2023provable}, Gumbel-Max~\cite{aaronson2023watermarking}, and SynthID~\cite{dathathri2024scalable}. Unlike \SMark, these methods do not provide identity-specific carrier separability across enrolled identities.

\textbf{Metrics.}
For substitution detection, we report TPR at a fixed FPR of 1\% (TPR@1\%), the corresponding accuracy at that threshold (ACC@1\%), and AUC. TPR@1\% is the primary metric because practical auditing requires very low false-alarm rates. For attribution, we report multi-class accuracy and macro-F1. For task performance, we report answer accuracy on Math, Pass@1/unit-test success on Code, and perplexity on General. For spoofing, we report attack success rate (ASR@1\%), the fraction of spoofed summaries incorrectly accepted as the claimed identity.

\textbf{Training configuration.}
We construct watermark carriers with green ratio $\rho=0.3$ and maximum overlap $\tau=0.3$. Adapters are trained on 5,000 samples with an 80/20 train-test split for one epoch using AdamW, learning rate $2\times 10^{-4}$, and temperature 0.7. Loss weights are $\lambda_{\mathrm{wm}}=\lambda_{\mathrm{sep}}=0.5$. We use a balanced 50\% mixture and provide sensitivity to substitution ratio in Appendix~\ref{app:hyper}. All experiments were conducted on one NVIDIA RTX 8000 GPU with 48GB memory.


\subsection{Identity-Substitution Detection}\label{sec:mainresults}

\begin{table}[t]
\centering
\scriptsize
\setlength{\tabcolsep}{2.0pt}
\renewcommand{\arraystretch}{0.9}
\caption{Summary of substitution detection across seven substitution pairs per dataset. We report average (AvgT) TPR@FPR=1\%, minimum (MinT) TPR@FPR=1\%, and average AUC. Full per-pair results are in Appendix~\ref{app:full_detection}. 
}
\label{tab:main_detection_summary}
\resizebox{0.65\columnwidth}{!}{%
\begin{tabular}{lccc ccc ccc}
\toprule
& \multicolumn{3}{c}{\textbf{Math}} 
& \multicolumn{3}{c}{\textbf{General}} 
& \multicolumn{3}{c}{\textbf{Code}} \\
\cmidrule(lr){2-4}\cmidrule(lr){5-7}\cmidrule(lr){8-10}
\textbf{Method} & \textbf{AvgT} & \textbf{MinT} & \textbf{AUC}
& \textbf{AvgT} & \textbf{MinT} & \textbf{AUC}
& \textbf{AvgT} & \textbf{MinT} & \textbf{AUC} \\
\midrule
BERT   & 0.7683 & 0.1232 & 0.9840 & 0.5439 & 0.0877 & 0.9401 & 0.6567 & 0.0000 & 0.9916 \\
GPT-2  & 0.3333 & 0.0000 & 0.7969 & 0.2506 & 0.0000 & 0.8101 & 0.3886 & 0.0000 & 0.8174 \\
T5     & 0.8676 & 0.6359 & 0.9787 & 0.5213 & 0.0000 & 0.9703 & 0.7494 & 0.1053 & 0.9975 \\
\SMark & \textbf{0.9994} & \textbf{0.9960} & \textbf{0.9996}
       & \textbf{0.9750} & \textbf{0.8740} & \textbf{0.9990}
       & \textbf{1.0000} & \textbf{1.0000} & \textbf{1.0000} \\
\bottomrule
\end{tabular}
}
\end{table}

Table~\ref{tab:main_detection_summary} summarizes substitution detection across Math, General, and Code. Each dataset aggregates the same seven substitution pairs: three cross-scale Qwen substitutions, two cross-family same-scale substitutions, and two joint family-and-scale substitutions. \SMark achieves consistently high TPR at the strict FPR$=1\%$ operating point, with average TPR@1\% of 0.9994 on Math, 0.975 on General, and 1.000 on Code. More importantly, its minimum TPR@1\% remains high across all pairs: 0.996 on Math, 0.874 on General, and 1.000 on Code.

The contrast matches the design goal of \SMark. Text classifiers implicitly learn domain- and dataset-specific regularities such as phrasing, formatting, and templates. These cues are strongest in more structured domains such as Code, where T5 achieves high AUC, but they are not identity-bound and can fail under closely related substitutions or domain changes. \SMark instead encodes identity-specific evidence into the summary channel through keyed vocabulary carriers and summary adapters, so substitution induces a predictable shift in summary-level token statistics rather than a fragile stylistic change. Full per-pair results, including the original Math, General, and Code tables, are reported in Appendix~\ref{app:full_detection}.

\begin{table*}[t]
\centering
\begin{minipage}[t]{0.49\textwidth}
\vspace{0pt}
\centering
\scriptsize
\setlength{\tabcolsep}{3.0pt}
\renewcommand{\arraystretch}{1.12}
\caption{Multi-class model identity attribution from reasoning summaries. }
\label{tab:attribution}
\resizebox{\linewidth}{!}{%
\begin{tabular}{@{}lcc cc cc@{}}
\toprule
& \multicolumn{2}{c}{\textbf{Math}} 
& \multicolumn{2}{c}{\textbf{Code}} 
& \multicolumn{2}{c}{\textbf{General}} \\
\cmidrule(lr){2-3}\cmidrule(lr){4-5}\cmidrule(lr){6-7}
\textbf{Method} 
& \textbf{ACC} & \textbf{F1} 
& \textbf{ACC} & \textbf{F1} 
& \textbf{ACC} & \textbf{F1} \\
\midrule
BERT   & 0.9386 & 0.9383 & 0.9580 & 0.9583 & 0.9201 & 0.9201 \\
GPT-2  & 0.7121 & 0.6981 & 0.5768 & 0.5754 & 0.7095 & 0.6894 \\
T5     & 0.9128 & 0.9122 & \textbf{0.9841} & \textbf{0.9841} & 0.9575 & 0.9575 \\
\SMark & \textbf{0.9958} & \textbf{0.9957} & 0.9678 & 0.9708 & \textbf{0.9718} & \textbf{0.9718} \\
\bottomrule
\end{tabular}
}
\end{minipage}
\hfill
\begin{minipage}[t]{0.49\textwidth}
\vspace{0pt}
\centering
\scriptsize
\setlength{\tabcolsep}{3.0pt}
\renewcommand{\arraystretch}{1.12}
\caption{Task performance with and without summary watermarking (WM). Higher is better for Acc and Pass@1; lower is better for PPL.}
\label{tab:answer_quality}
\resizebox{\linewidth}{!}{%
\begin{tabular}{@{}l cc cc cc@{}}
\toprule
& \multicolumn{2}{c}{\textbf{Math (Acc $\uparrow$)}} 
& \multicolumn{2}{c}{\textbf{General (PPL $\downarrow$)}} 
& \multicolumn{2}{c}{\textbf{Code (Pass@1 $\uparrow$)}} \\
\cmidrule(lr){2-3}\cmidrule(lr){4-5}\cmidrule(lr){6-7}
\textbf{Model} 
& \textbf{w/o WM} & \textbf{w/ WM} 
& \textbf{w/o WM} & \textbf{w/ WM} 
& \textbf{w/o WM} & \textbf{w/ WM} \\
\midrule
LLaMA-3B  & 0.2069 & \textbf{0.2159} & 8.1044  & \textbf{7.8450}  & 0.0290 & \textbf{0.0430} \\
Qwen-1.5B & \textbf{0.2971} & 0.2967 & 29.3089 & \textbf{25.6346} & \textbf{0.0170} & \textbf{0.0170} \\
Qwen-3B   & 0.4219 & \textbf{0.4240} & 10.1635 & \textbf{10.1429} & 0.0630 & \textbf{0.0760} \\
Qwen-7B   & 0.5034 & \textbf{0.5062} & 14.1778 & \textbf{13.8621} & \textbf{0.1920} & 0.1830 \\
\bottomrule
\end{tabular}
}
\end{minipage}
\end{table*}

\textbf{Attribution.}
Attribution uses the same standardized green-token statistic as detection, scoring the released summary against each enrolled identity and predicting the largest score. Table~\ref{tab:attribution} shows that \SMark supports multi-class attribution across datasets, reaching 0.9958/0.9957 ACC/F1 on Math and 0.9718/0.9718 on General. On Code, T5 slightly outperforms \SMark, reflecting the strong templatic structure of Code summaries. Nevertheless, \SMark remains competitive and does not rely on unkeyed stylistic cues.

\textbf{Task-facing answer quality.}
Table~\ref{tab:answer_quality} shows that summary-channel watermarking largely preserves final-answer quality. On Math and Code, performance is essentially unchanged after watermarking. On General, perplexity sometimes improves slightly, which we view as benign regularization of summary wording rather than improved reasoning. These results support localizing audit evidence to the summary channel rather than perturbing the task-facing answer.


\subsection{Generic Watermarks Are Not Enough}\label{sec:baselines}

\begin{table*}[t]
\centering
\scriptsize
\setlength{\tabcolsep}{2.12pt}
\renewcommand{\arraystretch}{1.12}
\caption{Substitution detection on Math with watermarking baselines. We report TPR@FPR=1\%, ACC@FPR=1\%, and AUC. Setting A$\rightarrow$B denotes serving B while claiming A.}
\label{tab:math_wm_baselines}
\resizebox{\textwidth}{!}{%
\begin{tabular}{@{}lcccccccccccc@{}}
\toprule
& \multicolumn{3}{c}{\textbf{Unigram}}
& \multicolumn{3}{c}{\textbf{Gumbel Max}}
& \multicolumn{3}{c}{\textbf{SynthID}}
& \multicolumn{3}{c}{\textbf{\SMark}} \\
\cmidrule(lr){2-4}
\cmidrule(lr){5-7}
\cmidrule(lr){8-10}
\cmidrule(lr){11-13}
\textbf{Setting}
& \textbf{TPR@1\%} & \textbf{ACC@1\%} & \textbf{AUC}
& \textbf{TPR@1\%} & \textbf{ACC@1\%} & \textbf{AUC}
& \textbf{TPR@1\%} & \textbf{ACC@1\%} & \textbf{AUC}
& \textbf{TPR@1\%} & \textbf{ACC@1\%} & \textbf{AUC} \\
\midrule

\rowcolor{black!8}
Qwen-7B$\rightarrow$Qwen-3B
& 0.4730 & 0.7315 & 0.9523
& 0.0050 & 0.4975 & 0.4828
& 0.0050 & 0.4975 & 0.5293
& \textbf{1.0000} & \textbf{0.9950} & \textbf{1.0000} \\

\rowcolor{black!8}
Qwen-7B$\rightarrow$Qwen-1.5B
& 0.0690 & 0.5295 & 0.7288
& 0.0300 & 0.5100 & 0.4968
& 0.0200 & 0.5050 & 0.5506
& \textbf{0.9960} & \textbf{0.9930} & \textbf{0.9974} \\

\rowcolor{black!8}
Qwen-3B$\rightarrow$Qwen-1.5B
& 0.0000 & 0.4950 & 0.2753
& 0.0950 & 0.5425 & 0.6150
& 0.0950 & 0.5425 & 0.6174
& \textbf{1.0000} & \textbf{0.9950} & \textbf{0.9998} \\

\rowcolor{black!16}
Qwen-3B$\rightarrow$LLaMA-3B
& 0.0010 & 0.4955 & 0.6302
& 0.0400 & 0.5150 & 0.4865
& 0.0800 & 0.5350 & 0.4961
& \textbf{1.0000} & \textbf{0.9950} & \textbf{0.9997} \\

\rowcolor{black!16}
LLaMA-3B$\rightarrow$Qwen-3B
& 0.1650 & 0.5775 & 0.7466
& 0.0000 & 0.4950 & 0.5585
& 0.0000 & 0.4950 & 0.5240
& \textbf{1.0000} & \textbf{0.9950} & \textbf{1.0000} \\

\rowcolor{black!22}
Qwen-7B$\rightarrow$LLaMA-3B
& 0.4510 & 0.7205 & 0.9214
& 0.1250 & 0.5575 & 0.6000
& 0.1000 & 0.5450 & 0.5987
& \textbf{1.0000} & \textbf{0.9950} & \textbf{1.0000} \\

\rowcolor{black!22}
LLaMA-3B$\rightarrow$Qwen-1.5B
& 0.0660 & 0.5280 & 0.6194
& 0.0000 & 0.4950 & 0.5471
& 0.0200 & 0.5050 & 0.5014
& \textbf{1.0000} & \textbf{0.9950} & \textbf{1.0000} \\

\bottomrule
\end{tabular}
}
\end{table*}

Table~\ref{tab:math_wm_baselines} compares \SMark with generic watermarking baselines under the same summary-channel substitution-detection protocol. 
Unlike \SMark, these baselines do not create identity-specific carriers. 
The results show that generic watermarking is insufficient for reliable identity auditing at the strict low-FPR operating point. 
Unigram shows some aggregate separability, e.g., AUC $=0.9523$ for Q7$\rightarrow$Q3 and $0.9214$ for Q7$\rightarrow$L3, but its TPR@1\% is only $0.4730$ and $0.4510$ in those cases and drops near zero elsewhere. Gumbel-Max and SynthID are weaker still.

By contrast, \SMark achieves TPR@1\% between $0.9960$ and $1.0000$, ACC@1\% between $0.9930$ and $0.9950$, and AUC between $0.9974$ and $1.0000$ across all seven substitution pairs. The key difference is that \SMark tests whether the summary matches the keyed footprint of the \emph{claimed identity}, not merely whether it contains a generic watermark. Identity-specific carrier construction gives enrolled identities separable token footprints, while adapter training amplifies the correct footprint and suppresses confusion with competitors. Appendix~\ref{app:ablations} further shows that Plain and shared-Greenlist variants fail at the strict FPR$=1\%$ operating point, confirming that the signal must be identity-specific and separable.


\subsection{Precision-Variant Diagnostics}\label{sec:exp:precision}

\begin{table*}[t]
\centering
\small
\setlength{\tabcolsep}{3.2pt}
\renewcommand{\arraystretch}{1.12}
\caption{Within-identity precision consistency across Math, Code, and General. FP16$\rightarrow$INT8 denotes serving a quantized variant while claiming the FP16 configuration of the same identity. Precision variants share the same carrier set and adapter; the verifier is calibrated on FP16 summaries.}
\label{tab:quantization}
\resizebox{0.9\textwidth}{!}{%
\begin{tabular}{@{}llccc ccc ccc@{}}
\toprule
& & \multicolumn{3}{c}{\textbf{Math}} 
& \multicolumn{3}{c}{\textbf{Code}} 
& \multicolumn{3}{c}{\textbf{General}} \\
\cmidrule(lr){3-5}\cmidrule(lr){6-8}\cmidrule(lr){9-11}
\textbf{Model} & \textbf{Setting}
& \textbf{TPR@1\%} & \textbf{ACC@1\%} & \textbf{AUC}
& \textbf{TPR@1\%} & \textbf{ACC@1\%} & \textbf{AUC}
& \textbf{TPR@1\%} & \textbf{ACC@1\%} & \textbf{AUC} \\
\midrule

\multirow{2}{*}{LLaMA-3B}
& FP16$\rightarrow$INT8
& 0.9990 & 0.9960 & 0.9990
& 0.9160 & 0.9530 & 0.9694
& 0.9580 & 0.9740 & 0.9971 \\
& FP16$\rightarrow$INT4
& 0.9990 & 0.9995 & 0.9990
& 0.9680 & 0.9790 & 0.9791
& 0.9930 & 0.9920 & 0.9998 \\

\midrule
\multirow{2}{*}{Qwen-1.5B}
& FP16$\rightarrow$INT8
& 0.1132 & 0.5521 & 0.7575
& 0.0360 & 0.5130 & 0.7010
& 0.4454 & 0.7182 & 0.8534 \\
& FP16$\rightarrow$INT4
& 0.3417 & 0.6663 & 0.9209
& 0.4120 & 0.7010 & 0.9497
& 0.7858 & 0.8884 & 0.9454 \\

\midrule
\multirow{2}{*}{Qwen-3B}
& FP16$\rightarrow$INT8
& 0.9860 & 0.9890 & 0.9973
& 0.9460 & 0.9680 & 0.9952
& 0.9910 & 0.9910 & 0.9964 \\
& FP16$\rightarrow$INT4
& 0.9960 & 0.9945 & 0.9986
& 0.9970 & 0.9935 & 0.9997
& 0.9950 & 0.9940 & 0.9989 \\

\midrule
\multirow{2}{*}{Qwen-7B}
& FP16$\rightarrow$INT8
& 0.9910 & 0.9920 & 0.9982
& 0.9910 & 0.9905 & 0.9996
& 0.9530 & 0.9715 & 0.9851 \\
& FP16$\rightarrow$INT4
& 0.9990 & 0.9975 & 0.9998
& 1.0000 & 0.9975 & 1.0000
& 0.9810 & 0.9875 & 0.9945 \\

\bottomrule
\end{tabular}
}
\end{table*}

Table~\ref{tab:quantization} evaluates \emph{within-identity precision consistency}: the provider serves an INT8 or INT4 quantized variant while claiming the FP16 model. Unlike cross-family or cross-scale substitution, precision variants share the same model identity, carrier set~$G_m$, and adapter~$\Delta_m$. We therefore treat this as a sidecar diagnostic rather than substitution detection over~$\mathcal M$: the verifier is calibrated on FP16 reference summaries and flags a quantized deployment only when its released summaries deviate from the FP16-calibrated footprint.

For LLaMA-3B, Qwen-3B, and Qwen-7B, both FP16$\rightarrow$INT8 and FP16$\rightarrow$INT4 are flagged reliably across Math, Code, and General, with AUCs mostly above $0.96$ and TPR@1\% often near $1.0$. INT4 is generally easier to flag than INT8, consistent with more aggressive quantization inducing a larger shift in the summary-channel footprint. The main exception is Qwen-1.5B: FP16$\rightarrow$INT8 remains close to the FP16 footprint, with TPR@1\% dropping to $0.1132$ on Math, $0.0360$ on Code, and $0.4454$ on General. INT4 is more detectable but still weaker than for larger identities. These results position FP16-calibrated precision consistency as a within-identity diagnostic that complements the cross-identity substitution detection studied above. Appendix~\ref{app:ablation_lqs}  ablates the precision-consistency regularizer introduced in Section~\ref{sec:lora}.


\subsection{Robustness and Boundary Analysis}\label{sec:exp:robustness}

\begin{wraptable}{r}{0.450\textwidth}
\vspace{-0.65in}
\centering
\scriptsize
\setlength{\tabcolsep}{2.4pt}
\renewcommand{\arraystretch}{0.9}
\caption{Robustness summary on Math across seven substitution pairs. For post-processing, we report average/min TPR@FPR=1\%; for spoofing, average/max ASR@FPR=1\%. Full results are in Appendix~\ref{app:robustness_full}. 
}
\label{tab:robustness_summary}
\resizebox{0.450\textwidth}{!}{%
\begin{tabular}{lcc}
\toprule
\textbf{Stress test} & \textbf{Avg.} & \textbf{Worst} \\
\midrule
Original TPR@1\% & 0.9994 & 0.9960 \\
Bounded edits TPR@1\% & 0.9992 & 0.9950 \\
Bilingual rewrite TPR@1\% & 0.7900 & 0.5740 \\
\midrule
Distilled spoofing ASR@1\% & 0.1651 & 0.6290 \\
Watermark-stealing ASR@1\% & 0.0417 & 0.2160 \\
\bottomrule
\end{tabular}
}
\end{wraptable}
Table~\ref{tab:robustness_summary} summarizes robustness under summary post-processing, rewriting, and spoofing. For bounded surface transformations, including sentence shuffle, random deletion, and word truncation, detection remains essentially unchanged: the minimum TPR@1\% across all edited summaries is 0.9950. This confirms that \SMark does not rely on exact phrasing or sentence order; its verifier aggregates summary-level token statistics that survive reordering and partial deletion.

Bilingual rewriting is stronger because it replaces tokens rather than rearranging them. It reduces average TPR@1\% to 0.7900 and worst-case TPR@1\% to 0.5740, showing that semantic-level rewriting can attenuate the lexical footprint. Thus, bounded edits are well tolerated, while rewriting that replaces many identity-discriminative tokens can reduce low-FPR detectability.

Spoofing gives a similar boundary-aware picture. Watermark-stealing is largely ineffective, with average ASR@1\% of 0.0417, suggesting that text-only token-statistic estimates cannot reproduce the keyed footprint. Distilled adapter spoofing is stronger: it is contained in several transfer directions but reaches worst-case ASR@1\% of 0.6290. We therefore treat spoofing robustness as empirical rather than guaranteed, and report full per-pair results, hardest transfer directions, and additional watermark removal stress tests in Appendix~\ref{app:robustness_full}.

\begin{tcolorbox}[
    enhanced, 
    width=\textwidth,
    arc=3mm, 
    boxrule=1pt, 
    colback=black!5, 
    colframe=black!75!black, 
    fonttitle=\bfseries, 
]
\noindent{\textbf{Takeaway.} Overall, \SMark turns reasoning summaries into a keyed, identity-specific audit signal: it enables reliable low-FPR detection and attribution from released summaries, outperforms shared and generic watermarking baselines, and preserves task-facing answer quality. The boundary analyses further clarify the scope of the guarantee: the signal is stable under bounded surface edits, useful but model-dependent for precision variants, and vulnerable to sufficiently strong rewriting or adaptive spoofing.}
\end{tcolorbox}

%% file: ack.tex
\section*{Acknowledgements} 
This work was supported in part by the National Science Foundation (NSF) grants 2541997, 2148309, 2315614 and 2337914, and the National Institutes of Health (NIH) grant R01HL184139.  Any opinions, findings, and conclusions or recommendations expressed in this material are those of the authors and do not necessarily reflect the views of the funding agencies.

%% file: appendix.tex
\section{Additional Related Work Discussions}

\subsection{LLM Watermarking}

LLM Watermarking has been widely studied as a way to embed verifiable provenance signals into generated text, enabling attribution, provenance tracking, or misuse detection while preserving output fluency~\cite{dathathri2024scalable}. 

Most existing approaches introduce controlled statistical biases at generation time, for example by favoring identity-specific token subsets or by shaping token selection patterns during decoding~\cite{kirchenbauer2023watermark,zhao2023provable}. Detection is typically performed through hypothesis testing on token frequencies or related likelihood statistics computed from the generated text.

A common design is to embed the watermark directly into task-facing outputs, usually the final answers returned to users. This makes the detectability of the watermark sensitive to task semantics and prompt distributions~\cite{piet2025markmywords}. In particular, low-entropy generations provide limited lexical degrees of freedom, so injecting a detectable bias can compete with answer fidelity and the watermark signal may be diluted by semantic constraints~\cite{liu2024survey}. Moreover, embedding signals into final answers raises concerns about potential degradation of answer quality and can be weakened by post-processing such as paraphrasing or rewriting~\cite{rastogi2024revisiting,jovanovic2024watermark,chen2024mark, Ghanim2025EvaluatingTR}. These limitations motivate moving audit evidence away from task answers and into a separate user-visible channel. Our work follows this direction by watermarking \emph{reasoning summaries} and enabling \emph{summary-only} verification, which reduces coupling to task constraints and preserves answer quality.

\subsection{Model Auditing}

Model auditing aims to verify whether a deployed machine learning system behaves as claimed, or whether specific properties of the model or inference process can be reliably certified~\cite{raji2020closing, casper2024black, sun2025invisible}. Prior work has explored a range of approaches, including behavioral testing through challenge–response protocols~\cite{he2018verideep,yang2025edgethemis}, hash tree-based verification~\cite{sun2025coin}, output-based fingerprinting using distributional or stylistic cues~\cite{gao2024model, pasquini2025llmmap}, and verification mechanisms that rely on access to internal model states, inference logs, or trusted execution environments~\cite{tramer2018slalom,chen2024zkml}. These methods have been applied to settings such as model integrity verification, inference correctness, compliance auditing~\cite{he2018verideep,tramer2018slalom}, and token count auditing~\cite{sun2025coin,wang2025predictive}. Beyond model-level verification, prior work has explored broader reliability and security challenges in machine learning systems across diverse deployment settings~\cite{jiang2024good,wei2026laua,zhang2026stealing,xu2026ask}.

Recent work has begun to explicitly study model substitution in LLM services, where a provider may serve responses generated by a different model than the one advertised~\cite{cai2025you}. This problem is particularly challenging in black-box settings, as modern language models with similar architectures or training objectives can exhibit highly overlapping behaviors. Behavioral differences between models from the same family or between different parameter scales may be subtle, task-dependent, and easy to smooth or obscure through prompt engineering or response post-processing~\cite{cai2025you, krishna2023paraphrasing}. As a result, auditing methods that rely solely on output behavior or stylistic features often provide limited robustness under realistic substitution strategies~\cite{cai2025you}.

Many existing auditing and verification techniques further assume privileged access that is unavailable in hosted LLM services~\cite{casper2024black}. For example, some approaches require access to internal activations, inference traces, or trusted instrumentation during model execution~\cite{tramer2018slalom, chen2024zkml,gao2024model}, while others rely on cooperative providers that accurately report model identity or generation metadata~\cite{cai2025you}. In contrast, end users and external auditors interacting with commercial LLM services typically observe only a fixed, user-facing interface and must perform verification post hoc, without control over prompts, sampling procedures, or internal logging~\cite{cai2025you, casper2024black}. 

Our work targets this more constrained but practically relevant auditing regime. We consider auditor-assisted model identity verification at deployment time, where a trusted auditor holds secret verification material but operates solely on user-visible outputs~\cite{adi2018turning}. By focusing on reasoning summaries as the sole audit surface, our approach departs from prior model auditing methods and addresses model substitution under an information regime that closely matches real-world hosted LLM services.

\section{Additional Experimental Details}
Here, we provide additional details on the experimental setup, including the datasets and models used, as well as details of the generation process.
\subsection{Datasets}
\label{app:datasets}
We use the following three datasets~\cite{wang2025predictive}.

\textbf{OpenR1-Math.} The Math dataset consists of multi-step mathematical reasoning problems requiring symbolic manipulation or logical derivation. Prompts are formatted to elicit a final answer together with a concise reasoning summary. Summaries in this dataset are typically structured and step-oriented, providing a high-density signal for evaluating watermark detectability under strong reasoning conditions.

\textbf{OpenR1-Code.} The Code dataset focuses on program synthesis and algorithmic problem solving. Each prompt requires generating code or an algorithmic solution, accompanied by a natural-language summary explaining the approach. Compared to Math, summaries exhibit greater lexical diversity and less rigid structure, testing robustness under heterogeneous reasoning styles.

\textbf{General Reasoning.} The General dataset includes open-domain reasoning and instruction-following tasks such as commonsense reasoning and explanation-style question answering. Summaries are shorter and more stylistically varied, representing the most challenging setting for summary-based auditing and reflecting common real-world usage.

\subsection{Models} 
Our primary benchmark uses four instruction-tuned checkpoints:
Qwen2.5-1.5B-Instruct, Qwen2.5-3B-Instruct,
Qwen2.5-7B-Instruct, and LLaMA-3.2-3B-Instruct.
These checkpoints are obtained from public repositories on Hugging Face.

The expanded evaluation in Appendix~\ref{app:full_detection} adds eight checkpoints, including 12 enrolled identities across four model families: Qwen (0.5B, 1.5B, 3B, 7B, 14B, and 32B), LLaMA (3B and 7B), Mistral (Ministral-3B and Mistral-7B), and Gemma (2B and 7B).
The primary benchmark and expanded evaluation use different training budgets, as specified in Section~\ref{sec:exp_setup} and Appendix~\ref{app:full_detection}, respectively.

\section{Watermark Design Details}
\subsection{Greenlist-based Watermarking}
We model a reasoning summary as a token sequence $s=(s_1,\dots,s_T)$ drawn from a vocabulary $V$. \textsc{SumMark} assigns each model identity $m\in\mathcal{M}$ a secret subset of tokens, called a \emph{greenlist} $G_m\subseteq V$, derived from the auditor's secret key. During summary generation, the watermark is implemented as an identity-dependent bias that increases the likelihood of sampling tokens from $G_m$ while preserving fluency and semantic fidelity.

At decoding step $t$, let $z_t\in\mathbb{R}^{|V|}$ be the output logits and $p_t(v)=\mathrm{softmax}(z_t)_v$ be the induced token distribution.
We quantify the instantaneous watermark signal via the log probability mass on $G_m$:
\begin{equation}
\log P(G_m\mid z_t)
\triangleq
\log \sum_{v\in G_m} e^{z_{t,v}}
-
\log \sum_{v\in V} e^{z_{t,v}}.
\label{eq:log_greenmass}
\end{equation}
Equivalently, $P(G_m\mid z_t)=\sum_{v\in G_m}p_t(v)$ represents the probability that the next generated token belongs to $G_m$.
Section~\ref{sec:lora} describes how identity-specific adapters are trained on supervised summary data to increase the quantity defined in Eq. \eqref{eq:log_greenmass}.

\subsection{Model Identities Capacity}
\label{app:bucket}

The parameters $\rho$ and $\tau$ constrain how many model identities can be supported for a fixed vocabulary. Let $V$ be the public vocabulary with size $n\triangleq |V|$. For each identity $m\in\mathcal{M}$, the greenlist is a subset $G_m\subseteq V$ with size $w\triangleq |G_m|$, and the green ratio constraint in Eq.~\eqref{eq:green_ratio} gives $w\approx \rho n$. To quantify collisions across identities, define the maximum pairwise intersection
\begin{equation}
    \lambda \triangleq \max_{m\neq m'} |G_m\cap G_{m'}|.
\end{equation}
Under the max-overlap constraint in Eq.~\eqref{eq:overlap}, we have $|G_m\cap G_{m'}|\le \tau |G_m|$ for all $m\neq m'$, hence $\lambda \le \tau w$. A standard packing bound for constant-size subsets then introduces
\begin{equation}
B \le \left\lfloor 1+\frac{w(n-w)}{w^2-n\lambda}\right\rfloor
\le
\left\lfloor 1+\frac{w(n-w)}{w^2-n\tau w}\right\rfloor,
\label{eq:capacity}
\end{equation}
where $B\triangleq|\mathcal{M}|$ is the number of identities and the bound is informative when the denominator is positive. Eq.~\eqref{eq:capacity} shows that smaller $\tau$ improves separability but can reduce capacity.

\section{Watermark Adapters Training}
\label{app:training}

To induce identity-specific token preferences in the \emph{summary channel}, we adopt a lightweight parameterization based on Low-Rank Adaptation (LoRA). For each identity $m\in\mathcal{M}$, we train an adapter $\Delta_m$ that biases the summary generator toward the identity-specific greenlist $G_m$. The base model parameters $\theta$ are frozen and only $\Delta_m$ is optimized, so the watermark is implemented as a small, identity-specific parameter add-on rather than a separate generation pipeline.

\subsection{Training Target}
We train each adapter $\Delta_m$ using supervised summary data. Each training example $(c, s^*) \sim \mathcal{D}$ consists of a conditioning input $c$ used to generate the summary and a reference summary $s^*=(w_1,\ldots,w_T)$. We use teacher forcing during training. At each decoding step, the model is given the ground-truth prefix of the summary and learns to predict the next token.

At step $t$, the adapted model $(\theta+\Delta_m)$ outputs a logit vector $z_t\in\mathbb{R}^{|V|}$ for every token in the vocabulary $V$. Applying a softmax converts these scores into a next-token probability distribution. Formally, we define
\begin{equation}
    p_t(v)=\mathrm{softmax}(z_t)_v,
\end{equation}
which denotes the probability of token $v$ at step $t$.

Each identity $m$ is associated with a predefined set of tokens, called the greenlist $G_m$. Intuitively, a watermark is stronger if the model tends to place more probability on these greenlist tokens during generation. We therefore measure the watermark strength by the total probability mass assigned to $G_m$:
\begin{equation}
    P(G_m|z_t)=\sum_{v \in G_m} p_t(v),
\end{equation}
which is the probability that the next token belongs to the greenlist. For numerical stability and convenience, we use its log-scaled form:
\begin{equation}
\log P(G_m \mid z_t)
=
\log \sum_{v \in G_m} e^{z_{t,v}}
\;-\;
\log \sum_{v \in V} e^{z_{t,v}}.
\end{equation}
Equivalently, $P(G_m \mid z_t) = \sum_{v \in G_m} p_t(v)$ is the probability that the next token sampled from $p_t$ lies in the greenlist $G_m$.
We refer to $P(G_m \mid z_t)$ as the \emph{green mass} at step $t$, and to $\log P(G_m \mid z_t)$ as its log-scaled form. A larger green mass means the model is more biased toward emitting identity-$m$ greenlist tokens, which results in a stronger and more detectable watermark signal.

During training, we freeze $\theta$ and update only $\Delta_m$. This ensures that the watermark is encoded through a small, identity-specific adapter, rather than by modifying or retraining the full model.

\subsection{Loss Components}

Given training pairs \((c,s^\ast)\sim\mathcal{D}\), we optimize the summary adapter \(\Delta_m\) with a base identity objective and an optional precision-consistency regularizer. The base identity objective is
\begin{equation}
L_{\mathrm{id}}
=
L_{\mathrm{sft}}
+
\lambda_{\mathrm{wm}}L_{\mathrm{wm}}
+
\lambda_{\mathrm{sep}}L_{\mathrm{sep}},
\end{equation}
where \(\lambda_{\mathrm{wm}},\lambda_{\mathrm{sep}}\ge 0\) control watermark strength and identity separability. Throughout this section, logits \(z_t\) are produced by the adapted model \(\theta_m+\Delta_m\) under teacher forcing on the reference summary. All losses are averaged over \((c,s^\ast)\sim\mathcal{D}\).

\paragraph{Utility constraint: \(L_{\mathrm{sft}}\).}
We use the standard teacher-forced negative log-likelihood on reference summaries:
\begin{equation}
L_{\mathrm{sft}}
=
-\mathbb{E}_{\mathcal{D}}
\log P_{\theta_m+\Delta_m}(s^\ast\mid c).
\end{equation}
This anchors generation to natural summaries and mitigates semantic drift caused by watermark pressure.

\paragraph{Detectability: \(L_{\mathrm{wm}}\).}
To strengthen the identity-specific watermark, we encourage larger average log probability mass on the carrier set \(G_m\):
\begin{equation}
L_{\mathrm{wm}}
=
-\mathbb{E}_{\mathcal{D}}
\left[
\frac{1}{T}
\sum_{t=1}^{T}
\log P(G_m\mid z_t)
\right].
\end{equation}
This term increases the statistical footprint associated with identity \(m\), making the watermark more detectable from generated summaries.

\paragraph{Attribution separability: \(L_{\mathrm{sep}}\).}
To reduce confusion between enrolled identities, we enforce that the carrier mass of the correct identity exceeds that of competing identities by a margin \(\mu\). Let \(\mathcal{N}\subset \mathcal{M}\setminus\{m\}\) be a sampled set of negative identities. We use a hardest-negative margin loss:
\begin{equation}
L_{\mathrm{sep}}
=
\mathbb{E}_{\mathcal{D}}
\left[
\frac{1}{T}
\sum_{t=1}^{T}
\max\left(
0,\,
\mu+
\max_{m'\in\mathcal{N}}
\log P(G_{m'}\mid z_t)
-
\log P(G_m\mid z_t)
\right)
\right].
\end{equation}
This term targets the most confusable competing identity and improves attribution reliability when carrier sets partially overlap.

\paragraph{Precision-consistency regularizer.}
For precision-variant diagnostics, FP16, INT8, and INT4 variants of the same base identity \(m\) share the same carrier set \(G_m\) and the same adapter \(\Delta_m\). We add a quantization-sensitive regularizer that encourages the FP16 execution to produce a stronger carrier-set footprint than quantized executions.

For a precision state \(p\in\{\mathrm{FP16},\mathrm{INT8},\mathrm{INT4}\}\), define the average carrier-set score
\begin{equation}
A_p(m)
=
\frac{1}{T}
\sum_{t=1}^{T}
\log P_p(G_m\mid z_t^{(p)}),
\end{equation}
where \(z_t^{(p)}\) denotes the logits produced by the shared adapter \(\Delta_m\) under precision state \(p\). We define
\begin{equation}
L_{\mathrm{qs}}
=
\mathbb{E}_{q\in\{\mathrm{INT8},\mathrm{INT4}\}}
\left[
\left(
\gamma -
\left(A_{\mathrm{FP16}}(m)-A_q(m)\right)
\right)_+
\right],
\end{equation}
where \((u)_+=\max(0,u)\) and \(\gamma>0\) is a precision-separation margin. This loss penalizes cases where the FP16 carrier-set score is not sufficiently larger than the INT8/INT4 score.

The full precision-sensitive training objective is
\begin{equation}
L
=
L_{\mathrm{id}}
+
\lambda_{\mathrm{qs}}L_{\mathrm{qs}}.
\end{equation}
This regularizer does not create precision-specific identities, carrier sets, or adapters. It only makes the shared identity-level verifier more informative for FP16-to-INT8/INT4 precision-consistency diagnostics.

\subsection{Detection}
\label{sec:detection}

At audit time, the auditor observes only the released summary $s$ and the claimed identity $m_c$.
Let $s=(x_1,\ldots,x_T)$ be the tokenized summary, where each token $x_t \in V$.
For each candidate identity $m\in M$ with keyed carrier bucket $G_m \subseteq V$, define
\[
g_m(x_t) \triangleq \mathbb{I}[x_t \in G_m],
\qquad
C_m(s) \triangleq \sum_{t=1}^{T} g_m(x_t).
\]
Although a binomial null would suggest $\mathbb{E}[C_m(s)]\approx \rho T$, all reported decisions use empirical calibration rather than this approximation. For each enrolled identity $m$, we estimate length-conditioned calibration statistics $\mu_m(T)$ and $\sigma_m(T)$ from held-out benign summaries generated under the enrolled identity and its summary adapter. The calibrated score is
\[
Z_m(s) \triangleq \frac{C_m(s)-\mu_m(T)}{\sigma_m(T)+\epsilon},
\]
where $\epsilon$ is a small numerical constant. Intuitively, summaries consistent with identity $m$ should have calibrated scores comparable to the held-out benign calibration distribution, while summaries generated under substituted identities tend to have lower scores for the claimed carrier bucket.

\paragraph{Model substitution detection.}
Given a claimed identity $m_c$, the verifier accepts the claim if
\[
Z_{m_c}(s) \ge \eta_{m_c}(\alpha),
\]
where $\eta_{m_c}(\alpha)$ is chosen as the lower-tail empirical threshold on held-out benign summaries so that the false-positive rate is at most $\alpha$. Otherwise, the verifier flags the released summary as inconsistent with the claimed identity. Thus the operating threshold is calibrated from held-out summaries rather than assumed from the binomial approximation.

\paragraph{Model attribution.}
For attribution, the auditor scores the released summary against every enrolled identity and predicts
\[
\widehat{m}(s) \triangleq \arg\max_{m\in M} Z_m(s),
\]
optionally abstaining when the top-two score margin is below a fixed threshold. Both detection and attribution rely only on summary-level token statistics and the keyed carrier buckets.

\begin{table*}[t]
\centering
\small
\caption{Substitution detection on \textbf{Math} (TPR@FPR=1\%, ACC@FPR=1\%, AUC).
Setting $A\!\to\!B$: serve $B$, claim $A$. Bold is best.}
\setlength{\tabcolsep}{3.2pt}
\renewcommand{\arraystretch}{1.08}
\resizebox{\linewidth}{!}{%
\begin{tabular}{lccc ccc ccc ccc}
\toprule
 &
\multicolumn{3}{c}{\textbf{BERT}} &
\multicolumn{3}{c}{\textbf{GPT-2}} &
\multicolumn{3}{c}{\textbf{T5}} &
\multicolumn{3}{c}{\textbf{\SMark}} \\
\cmidrule(lr){2-4}\cmidrule(lr){5-7}\cmidrule(lr){8-10}\cmidrule(lr){11-13}
\textbf{Setting}
& \textbf{TPR@1\%} & \textbf{ACC@1\%} & \textbf{AUC}
& \textbf{TPR@1\%} & \textbf{ACC@1\%} & \textbf{AUC}
& \textbf{TPR@1\%} & \textbf{ACC@1\%} & \textbf{AUC}
& \textbf{TPR@1\%} & \textbf{ACC@1\%} & \textbf{AUC} \\
\midrule

\rowcolor{black!7}
Qwen-7B$\to$Qwen-3B
& 0.1232 & 0.5602 & 0.9221
& 0.0000 & 0.5000 & 0.4458
& 0.6359 & 0.8053 & 0.9348
& \textbf{1.0000} & \textbf{0.9950} & \textbf{1.0000} \\

\rowcolor{black!8}
Qwen-7B$\to$Qwen-1.5B
& 0.9412 & 0.9706 & 0.9955
& 0.2105 & 0.5965 & 0.6037
& 0.9804 & 0.9902 & 0.9939
& \textbf{0.9960} & \textbf{0.9930} & \textbf{0.9974} \\

\rowcolor{black!8}
Qwen-3B$\to$Qwen-1.5B
& 0.9720 & 0.9790 & 0.9959
& 0.2807 & 0.6316 & 0.8339
& 0.9804 & 0.9748 & 0.9976
& \textbf{1.0000} & \textbf{0.9950} & \textbf{0.9998} \\

\rowcolor{black!16}
Qwen-3B$\to$LLaMA-3B
& 0.9076 & 0.9468 & 0.9939
& 0.6842 & 0.8333 & 0.9120
& 0.7003 & 0.8235 & 0.9605
& \textbf{1.0000} & \textbf{0.9950} & \textbf{0.9997} \\

\rowcolor{black!16}
LLaMA-3B$\to$Qwen-3B
& 0.8431 & 0.9202 & 0.9939
& 0.3684 & 0.6754 & 0.9477
& 0.8627 & 0.9300 & 0.9745
& \textbf{1.0000} & \textbf{0.9950} & \textbf{1.0000} \\

\rowcolor{black!22}
Qwen-7B$\to$LLaMA-3B
& 0.6050 & 0.8011 & 0.9871
& 0.2281 & 0.6053 & 0.8507
& 0.9272 & 0.9510 & 0.9941
& \textbf{1.0000} & \textbf{0.9950} & \textbf{1.0000} \\

\rowcolor{black!22}
LLaMA-3B$\to$Qwen-1.5B
& 0.9860 & 0.9930 & 0.9999
& 0.5614 & 0.7719 & 0.9843
& 0.9860 & 0.9930 & 0.9955
& \textbf{1.0000} & \textbf{0.9950} & \textbf{1.0000} \\

\bottomrule
\end{tabular}%
}
\vspace{-0.1in}
\label{tab:sub_math}
\end{table*}

\begin{table*}[t]
\centering
\small
\caption{Substitution detection on \textbf{General} (TPR@FPR=1\%, ACC@FPR=1\%, AUC). Setting $A\!\to\!B$: serve $B$, claim $A$. Q7/Q3/Q1.5 = Qwen-7B/3B/1.5B, L3 = LLaMA-3B. Bold is best.}
\setlength{\tabcolsep}{3.2pt}
\renewcommand{\arraystretch}{1.08}
\resizebox{\linewidth}{!}{%
\begin{tabular}{lccc ccc ccc ccc}
\toprule
 &
\multicolumn{3}{c}{\textbf{BERT}} &
\multicolumn{3}{c}{\textbf{GPT-2}} &
\multicolumn{3}{c}{\textbf{T5}} &
\multicolumn{3}{c}{\textbf{\SMark}} \\
\cmidrule(lr){2-4}\cmidrule(lr){5-7}\cmidrule(lr){8-10}\cmidrule(lr){11-13}
\textbf{Setting} & \textbf{TPR@1\%} & \textbf{ACC@1\%} & \textbf{AUC}
& \textbf{TPR@1\%} & \textbf{ACC@1\%} & \textbf{AUC}
& \textbf{TPR@1\%} & \textbf{ACC@1\%} & \textbf{AUC}
& \textbf{TPR@1\%} & \textbf{ACC@1\%} & \textbf{AUC} \\
\midrule

\rowcolor{black!8}
Qwen-7B$\to$Qwen-3B
& 0.7895 & 0.8772 & 0.9621
& 0.0000 & 0.5000 & 0.6331
& 0.7193 & 0.8596 & 0.9898
& \textbf{0.9990} & \textbf{0.9945} & \textbf{0.9999} \\

\rowcolor{black!8}
Qwen-7B$\to$Qwen-1.5B
& 0.1228 & 0.5439 & 0.8689
& 0.0526 & 0.5263 & 0.5389
& 0.0000 & 0.5000 & 0.9514
& \textbf{0.8740} & \textbf{0.9320} & \textbf{0.9954} \\

\rowcolor{black!8}
Qwen-3B$\to$Qwen-1.5B
& 0.0877 & 0.5439 & 0.9267
& 0.1053 & 0.5526 & 0.7014
& 0.0000 & 0.5000 & 0.9338
& \textbf{1.0000} & \textbf{0.9950} & \textbf{1.0000} \\

\rowcolor{black!16}
Qwen-3B$\to$LLaMA-3B
& 0.5439 & 0.7719 & 0.9298
& 0.3509 & 0.6754 & 0.9464
& 0.5088 & 0.7544 & \textbf{1.0000}
& \textbf{1.0000} & \textbf{0.9950} & 0.9999 \\

\rowcolor{black!16}
LLaMA-3B$\to$Qwen-3B
& 0.8596 & 0.9211 & 0.9917
& 0.3684 & 0.6842 & 0.9372
& 0.9474 & 0.9561 & 0.9800
& \textbf{1.0000} & \textbf{0.9950} & \textbf{1.0000} \\

\rowcolor{black!22}
Qwen-7B$\to$LLaMA-3B
& 0.5965 & 0.7807 & 0.9231
& 0.3509 & 0.6754 & 0.9323
& 0.5439 & 0.7719 & 0.9471
& \textbf{0.9920} & \textbf{0.9910} & \textbf{0.9988} \\

\rowcolor{black!22}
LLaMA-3B$\to$Qwen-1.5B
& 0.8070 & 0.8947 & 0.9785
& 0.5263 & 0.7632 & 0.9812
& 0.9298 & 0.9474 & 0.9898
& \textbf{0.9600} & \textbf{0.9750} & \textbf{0.9988} \\

\bottomrule
\end{tabular}%
}
\label{tab:subs_general}
\end{table*}

\begin{table*}[t]
\centering
\small
\caption{Substitution detection on \textbf{Code} (TPR@FPR=1\%, ACC@FPR=1\%, AUC). Setting $A\!\to\!B$: serve $B$, claim $A$. Bold is best.}
\setlength{\tabcolsep}{3.2pt}
\renewcommand{\arraystretch}{1.08}
\resizebox{\textwidth}{!}{%
\begin{tabular}{lccc ccc ccc ccc}
\toprule
 &
\multicolumn{3}{c}{\textbf{BERT}} &
\multicolumn{3}{c}{\textbf{GPT-2}} &
\multicolumn{3}{c}{\textbf{T5}} &
\multicolumn{3}{c}{\textbf{\SMark}} \\
\cmidrule(lr){2-4}\cmidrule(lr){5-7}\cmidrule(lr){8-10}\cmidrule(lr){11-13}
\textbf{Setting}& \textbf{TPR@1\%} & \textbf{ACC@1\%} & \textbf{AUC}
& \textbf{TPR@1\%} & \textbf{ACC@1\%} & \textbf{AUC}
& \textbf{TPR@1\%} & \textbf{ACC@1\%} & \textbf{AUC}
& \textbf{TPR@1\%} & \textbf{ACC@1\%} & \textbf{AUC} \\
\midrule

\rowcolor{black!8}
Qwen-7B$\to$Qwen-3B
& 0.6491 & 0.8246 & 0.9825
& 0.0000 & 0.5000 & 0.5720
& 0.8947 & 0.9474 & 0.9963
& \textbf{1.0000} & \textbf{0.9950} & \textbf{1.0000} \\

\rowcolor{black!8}
Qwen-7B$\to$Qwen-1.5B
& 0.8246 & 0.9123 & 0.9834
& 0.1400 & 0.5700 & 0.6690
& \textbf{1.0000} & \textbf{1.0000} & \textbf{1.0000}
& \textbf{1.0000} & 0.9950 & \textbf{1.0000} \\

\rowcolor{black!8}
Qwen-3B$\to$Qwen-1.5B
& 0.0000 & 0.5000 & 0.9769
& 0.5090 & 0.7110 & 0.7960
& 0.1053 & 0.5526 & 0.9982
& \textbf{1.0000} & \textbf{0.9950} & \textbf{1.0000} \\

\rowcolor{black!16}
Qwen-3B$\to$LLaMA-3B
& 0.2281 & 0.6140 & 0.9991
& 0.7370 & 0.8250 & 0.9330
& 0.3684 & 0.6842 & 0.9942
& \textbf{1.0000} & \textbf{0.9950} & \textbf{1.0000} \\

\rowcolor{black!16}
LLaMA-3B$\to$Qwen-3B
& 0.9474 & 0.9737 & 0.9991
& 0.4040 & 0.6750 & 0.9160
& 0.8947 & 0.9474 & 0.9935
& \textbf{1.0000} & \textbf{0.9950} & \textbf{1.0000} \\

\rowcolor{black!22}
Qwen-7B$\to$LLaMA-3B
& 0.9825 & 0.9912 & \textbf{1.0000}
& 0.2810 & 0.6400 & 0.8860
& \textbf{1.0000} & \textbf{1.0000} & \textbf{1.0000}
& \textbf{1.0000} & 0.9950 & \textbf{1.0000} \\

\rowcolor{black!22}
LLaMA-3B$\to$Qwen-1.5B
& 0.9649 & 0.9825 & \textbf{1.0000}
& 0.6490 & 0.7980 & 0.9500
& 0.9825 & 0.9912 & \textbf{1.0000}
& \textbf{1.0000} & \textbf{0.9950} & \textbf{1.0000} \\

\bottomrule
\end{tabular}%
}
\vspace{-0.1in}
\label{tab:sub_code}
\end{table*}
\section{Additional Experiments}

\subsection{Identity-Substitution Detection}
\textbf{Per-pair Substitution Detection Results.}
\label{app:full_detection}
Tables~\ref{tab:sub_math}, \ref{tab:subs_general}, and \ref{tab:sub_code} report model substitution detection on Math, General, and Code, respectively. The first three rows in each table correspond to \emph{within-family size substitution} (Qwen $\to$ Qwen), the next two to \emph{cross-family substitution} (Qwen $\leftrightarrow$ LLaMA), and the last two to \emph{combined} substitutions (family + size).

Across datasets and substitution types, \SMark is consistently strong under strict auditing constraints, achieving very high TPR@FPR$=1\%$ in most settings and uniformly high AUC. In contrast, text classifiers (BERT/GPT-2/T5) are \emph{unstable across settings}: they can show high AUC in some pairs, yet their TPR@1\% varies widely, which is most notably for within-family size substitutions where stylistic cues are subtle and easily washed out. This AUC-TPR gap indicates that while some substitutions are separable in aggregate, surface cues are often too brittle to support reliable decisions at a strict operating point (FPR$=1\%$). For instance on Math, BERT’s TPR@1\% collapses for Qwen-7B$\to$Qwen-3B (0.1232) while \SMark remains 1.0000, illustrating that style-based signals are not reliable when the substituted model is closely related to the claimed one.

This contrast matches the design intent of \SMark. Text classifiers implicitly learn domain- and dataset-specific surface regularities (phrasing, formatting, recurring templates), which may transfer poorly across tasks and are weakest when two models share the same family and training pipeline. By comparison, \emph{\SMark encodes {identity-specific} evidence into the summary channel via keyed vocabulary buckets and watermark adapters, so substitution induces a predictable shift in summary-level token statistics rather than a fragile stylistic change.} This explains why \SMark remains effective for size substitution, family substitution, and combined substitution across domains. 

\textit{Why General can be harder.}
We see a few weaker cases on General dataset (Table~\ref{tab:subs_general}); for example, Qwen-7B$\to$Qwen-1.5B has TPR@1\% = 0.8740. General summaries are more conversational and stylistically diverse, with fewer repeated templates than Code, which increases lexical entropy and reduces the effective per-token evidence that our count-based detector can accumulate. Moreover, General summaries rely heavily on high-frequency function words and generic phrasing; when ubiquitous tokens fall into a greenlist, non-watermarked summaries can match portions of the greenlist by chance, narrowing separation specifically at very low-FPR thresholds. Importantly, separability remains strong overall: this same setting retains a high AUC (0.9954), indicating the watermark signal is present but stricter operating points (e.g., FPR$=1\%$) can incur more misses.

\textit{Why Code can make baselines look strong.}
On the Code dataset (Table~\ref{tab:sub_code}), summaries are more structured and often follow recurring templates, e.g., step-by-step debugging narratives, standard error-handling language, and repeated references to functions/variables. This regularity can make strong encoders (notably T5) appear competitive in some substitution pairs by exploiting surface-level cues. However, this advantage is largely \emph{domain-specific}: the same classifiers are much less consistent on Math and General, reinforcing that their performance often reflects exploitable formatting/style regularities rather than a verifiable, identity-bound signal. By contrast, \SMark remains reliable across domains and substitution types because its evidence is keyed and identity-specific rather than stylistic.

\begin{table*}[t]
\centering
\small
\caption{Substitution detection on \textbf{Math} (TPR@FPR=1\%, ACC@FPR=1\%, AUC).
Setting $A\!\to\!B$: serve $B$, claim $A$.
Q7/Q3/Q1.5 = Qwen-7B/3B/1.5B, L3 = LLaMA-3B. Bold is best.}
\setlength{\tabcolsep}{3.0pt}
\renewcommand{\arraystretch}{1.08}
\resizebox{\linewidth}{!}{%
\begin{tabular}{lccc ccc ccc ccc ccc}
\toprule
 &
\multicolumn{3}{c}{\textbf{BERT}} &
\multicolumn{3}{c}{\textbf{GPT-2}} &
\multicolumn{3}{c}{\textbf{T5}} &
\multicolumn{3}{c}{\textbf{IDP}} &
\multicolumn{3}{c}{\textbf{\SMark}} \\
\cmidrule(lr){2-4}\cmidrule(lr){5-7}\cmidrule(lr){8-10}\cmidrule(lr){11-13}\cmidrule(lr){14-16}
\textbf{Setting}
& \textbf{TPR@1\%} & \textbf{ACC@1\%} & \textbf{AUC}
& \textbf{TPR@1\%} & \textbf{ACC@1\%} & \textbf{AUC}
& \textbf{TPR@1\%} & \textbf{ACC@1\%} & \textbf{AUC}
& \textbf{TPR@1\%} & \textbf{ACC@1\%} & \textbf{AUC}
& \textbf{TPR@1\%} & \textbf{ACC@1\%} & \textbf{AUC} \\
\midrule

\rowcolor{black!7}
$Q7\to Q3$ & 0.1232 & 0.5602 & 0.9221 & 0.0000 & 0.5000 & 0.4458 & 0.6359 & 0.8053 & 0.9348 & 0.0000 & 0.5000 & 0.5000 & \textbf{1.0000} & \textbf{0.9950} & \textbf{1.0000} \\

\rowcolor{black!8}
$Q7\to Q1.5$ & 0.9412 & 0.9706 & 0.9955 & 0.2105 & 0.5965 & 0.6037 & 0.9804 & 0.9902 & 0.9939 & 0.0000 & 0.5000 & 0.5000 & \textbf{0.9960} & \textbf{0.9930} & \textbf{0.9974} \\

\rowcolor{black!8}
$Q3\to Q1.5$ & 0.9720 & 0.9790 & 0.9959 & 0.2807 & 0.6316 & 0.8339 & 0.9804 & 0.9748 & 0.9976 & \textbf{1.0000} & \textbf{1.0000} & \textbf{1.0000} & \textbf{1.0000} & 0.9950 & 0.9998 \\

\rowcolor{black!16}
$Q3\to L3$ & 0.9076 & 0.9468 & 0.9939 & 0.6842 & 0.8333 & 0.9120 & 0.7003 & 0.8235 & 0.9605 & 0.0000 & 0.5000 & 0.5000 & \textbf{1.0000} & \textbf{0.9950} & \textbf{0.9997} \\

\rowcolor{black!16}
$L3\to Q3$ & 0.8431 & 0.9202 & 0.9939 & 0.3684 & 0.6754 & 0.9477 & 0.8627 & 0.9300 & 0.9745 & 0.0000 & 0.5000 & 0.5000 & \textbf{1.0000} & \textbf{0.9950} & \textbf{1.0000} \\

\rowcolor{black!22}
$Q7\to L3$ & 0.6050 & 0.8011 & 0.9871 & 0.2281 & 0.6053 & 0.8507 & 0.9272 & 0.9510 & 0.9941 & 0.0000 & 0.5000 & 0.5000 & \textbf{1.0000} & \textbf{0.9950} & \textbf{1.0000} \\

\rowcolor{black!22}
$L3\to Q1.5$ & 0.9860 & 0.9930 & 0.9999 & 0.5614 & 0.7719 & 0.9843 & 0.9860 & 0.9930 & 0.9955 & \textbf{1.0000} & \textbf{1.0000} & \textbf{1.0000} & \textbf{1.0000} & 0.9950 & \textbf{1.0000} \\

\bottomrule
\end{tabular}%
}
\label{tab:sub_math_idp}
\end{table*}

\begin{table}[t]
\centering
\small
\caption{Attribution performance on \textbf{Math}. Higher is better.}
\label{tab:attr_math_idp}
\setlength{\tabcolsep}{6pt}
\renewcommand{\arraystretch}{1.1}
\begin{tabular}{lcc}
\toprule
\textbf{Method} & \textbf{ACC} & \textbf{Macro-F1} \\
\midrule
BERT & 0.9386 & 0.9383 \\
GPT-2 & 0.7121 & 0.6981 \\
T5   & 0.9128 & 0.9122 \\
IDP  & 0.7500 & 0.6670 \\
\textbf{\SMark} & \textbf{0.9958} & \textbf{0.9957} \\
\bottomrule
\end{tabular}
\end{table}

\begin{table*}[t]
\centering
\small
\setlength{\tabcolsep}{3.2pt}
\renewcommand{\arraystretch}{1.12}
\caption{Identity-Substitution detection on Math. SynthID and Gumbel Max are applied directly to the task-facing answer, while \SMark applies the watermark only to the summary channel. We report TPR@FPR=1\%, ACC@FPR=1\%, and AUC.}
\label{tab:math_substitution_detection}
\resizebox{0.95\textwidth}{!}{%
\begin{tabular}{@{}lccc ccc ccc@{}}
\toprule
& \multicolumn{3}{c}{\textbf{SynthID}}
& \multicolumn{3}{c}{\textbf{Gumbel Max}}
& \multicolumn{3}{c}{\textbf{\SMark}} \\
\cmidrule(lr){2-4}\cmidrule(lr){5-7}\cmidrule(lr){8-10}
\textbf{Setting}
& \textbf{TPR@1\%} & \textbf{ACC@1\%} & \textbf{AUC}
& \textbf{TPR@1\%} & \textbf{ACC@1\%} & \textbf{AUC}
& \textbf{TPR@1\%} & \textbf{ACC@1\%} & \textbf{AUC} \\
\midrule

\rowcolor{black!8}
Q7$\rightarrow$ Q3 & 0.0000 & 0.4950 & 0.5350 & 0.0500 & 0.5200 & 0.6041 & \textbf{1.0000} & \textbf{0.9950} & \textbf{1.0000} \\

\rowcolor{black!8}
Q7$\rightarrow$ Q1.5 & 0.0350 & 0.5125 & 0.5623 & 0.0300 & 0.5100 & 0.5463 & \textbf{0.9960} & \textbf{0.9930} & \textbf{0.9974} \\

\rowcolor{black!8}
Q3$\rightarrow$ Q1.5 & 0.0700 & 0.5300 & 0.5235 & 0.0200 & 0.5050 & 0.4112 & \textbf{1.0000} & \textbf{0.9950} & \textbf{0.9998} \\

\rowcolor{black!16}
Q3$\rightarrow$ L3 & 0.0250 & 0.5075 & 0.5216 & 0.0450 & 0.5175 & 0.4143 & \textbf{1.0000} & \textbf{0.9950} & \textbf{0.9997} \\

\rowcolor{black!16}
L3$\rightarrow$ Q3 & 0.0000 & 0.4950 & 0.4727 & 0.0000 & 0.4950 & 0.4770 & \textbf{1.0000} & \textbf{0.9950} & \textbf{1.0000} \\

\rowcolor{black!22}
Q7$\rightarrow$ L3 & 0.0750 & 0.5325 & 0.5737 & 0.0150 & 0.5025 & 0.4531 & \textbf{1.0000} & \textbf{0.9950} & \textbf{1.0000} \\

\rowcolor{black!22}
L3$\rightarrow$ Q1.5 & 0.0250 & 0.5075 & 0.4680 & 0.0100 & 0.5000 & 0.4352 & \textbf{1.0000} & \textbf{0.9950} & \textbf{1.0000} \\

\bottomrule
\end{tabular}
}
\vspace{-0.1in}
\end{table*}

\textbf{Comparing with Identity prompting (IDP).}
Prompt the model to disclose its identity (e.g., “Who are you?”, “Who developed you?”) and string-match against known identifiers (family, developer), following the simple classifier setup in Huang et al.~\cite{huang2025exploring} We compare our method \SMark  with IDP on Math Dataset. Table~\ref{tab:sub_math_idp} shows that IDP performs poorly on substitution detection across most settings on the Math dataset. In many cases, IDP yields a TPR@1\% of 0 with ACC@1\% fixed at 0.5 and AUC at 0.5, which is equivalent to random guessing. This indicates that identity prompting fails to distinguish between the claimed model and the serving model under strict false positive constraints. Only in a few settings does IDP reach perfect scores, but these results are inconsistent and highly setting-dependent. Overall, the results suggest that IDP lacks robustness and cannot reliably detect model substitution in this task.

And we also conduct attribution detection for IDP. Table~\ref{tab:attr_math_idp} reports attribution performance on the Math dataset. The IDP baseline achieves relatively low performance on the Math dataset, with an accuracy of 0.75 and a Macro-F1 of 0.667. This gap compared to stronger attribution methods indicates that identity prompting combined with simple string matching is not sufficient for robust attribution.

\textbf{Comparing with answer-level watermarking methods.} Table~\ref{tab:math_substitution_detection} shows that applying distortion-free watermarks directly to task-facing answers is insufficient for identity substitution detection on Math. Both SynthID and Gumbel Max achieve near-zero TPR@FPR=1\% in most settings, with accuracies close to random guessing and AUCs mostly near chance. In contrast, \SMark achieves 0.996–1.000 TPR@1\%, 0.993–0.995 ACC@1\%, and consistently high AUC across all substitution pairs. This gap indicates that the task-facing answer is a weak carrier for identity auditing under strict false-alarm constraints.

\begin{table*}[t]
\centering
\small
\caption{Identity-Substitution detection performance on \textbf{Math} dataset.
We report TPR@FPR=1\% and TPR@FPR=5\%.
Best is bold.}
\setlength{\tabcolsep}{4.2pt}
\renewcommand{\arraystretch}{1.08}
\resizebox{\linewidth}{!}{%
\begin{tabular}{l cc cc cc cc}
\toprule
& \multicolumn{2}{c}{\textbf{BERT}}
& \multicolumn{2}{c}{\textbf{GPT-2}}
& \multicolumn{2}{c}{\textbf{T5}}
& \multicolumn{2}{c}{\textbf{\SMark}} \\
\cmidrule(lr){2-3}\cmidrule(lr){4-5}\cmidrule(lr){6-7}\cmidrule(lr){8-9}
\textbf{Setting}
& \textbf{TPR@1\%} & \textbf{TPR@5\%}
& \textbf{TPR@1\%} & \textbf{TPR@5\%}
& \textbf{TPR@1\%} & \textbf{TPR@5\%}
& \textbf{TPR@1\%} & \textbf{TPR@5\%} \\
\midrule

\rowcolor{black!7}
Qwen-7B$\to$Qwen-3B
& 0.1232 & 0.5322
& 0.0000 & 0.0351
& 0.6359 & 0.8095
& \textbf{1.0000} & \textbf{1.0000} \\

\rowcolor{black!8}
Qwen-7B$\to$Qwen-1.5B
& 0.9412 & 0.9720
& 0.2105 & 0.2807
& 0.9804 & 0.9804
& \textbf{0.9960} & \textbf{0.9970} \\

\rowcolor{black!8}
Qwen-3B$\to$Qwen-1.5B
& 0.9720 & 0.9720
& 0.2807 & 0.2807
& 0.9804 & 0.9804
& \textbf{1.0000} & \textbf{1.0000} \\

\rowcolor{black!16}
Qwen-3B$\to$LLaMA-3B
& 0.9076 & 0.9692
& 0.6842 & 0.8070
& 0.7003 & 0.7199
& \textbf{1.0000} & \textbf{1.0000} \\

\rowcolor{black!16}
LLaMA-3B$\to$Qwen-3B
& 0.8431 & 0.9160
& 0.3684 & 0.7368
& 0.8627 & 0.8627
& \textbf{1.0000} & \textbf{1.0000} \\

\rowcolor{black!22}
Qwen-7B$\to$LLaMA-3B
& 0.6050 & 0.9160
& 0.2281 & 0.4737
& 0.9272 & 0.9888
& \textbf{1.0000} & \textbf{1.0000} \\

\rowcolor{black!22}
LLaMA-3B$\to$Qwen-1.5B
& 0.9860 & 0.9860
& 0.5614 & 0.9474
& 0.9860 & 0.9860
& \textbf{1.0000} & \textbf{1.0000} \\

\bottomrule
\end{tabular}%
}
\label{tab:math_tpr_compare}
\end{table*}

\textbf{Substitution results on TPR@FPR = 5\%. }
Table~\ref{tab:math_tpr_compare} compares our method with text-only baselines under a relaxed auditing constraint of FPR = 5\% on Math dataset. Our approach achieves perfect or near-perfect detection across all substitution settings, consistently reaching TPR near to 1. In contrast, the baselines improve or remain unchanged under the relaxed threshold, but remain less consistent across model pairs. BERT improves under the relaxed threshold but remains unstable, particularly for cross-family substitutions. GPT-2 performs poorly in most settings, indicating limited sensitivity to model-identity cues. T5 shows relatively strong performance on some size substitutions but degrades noticeably on harder cross-family cases. These results demonstrate that even when allowing higher false positive rates, our method provides significantly stronger and more reliable substitution detection than text-based classifiers, highlighting its robustness and scalability under practical auditing conditions.

\textbf{Substitution Detection under Quantized Serving.}
Table~\ref{tab:precision_robustness} evaluates whether substitution detection remains reliable when the served substitute model is run at different numerical precisions. Across FP16, INT8, and INT4, \SMark maintains consistently high performance: TPR@1\% remains between 0.9860 and 1.0000, ACC@1\% stays around 0.9880--0.9950, and AUC is close to 1.0 in all settings.

The results show that quantizing the substituted model does not weaken the identity-substitution signal. In most cases, INT8 and INT4 perform comparably to FP16, and sometimes even slightly better. This suggests that \SMark's detector is primarily driven by the mismatch between the claimed identity's summary-channel footprint and the served model's footprint, rather than by fine-grained numerical precision effects.

\textbf{Expanded enrolled universe.}
To study scalability, we enlarge the enrolled universe to 12 identities from four families: Qwen (0.5B, 1.5B, 3B, 7B, 14B, 32B), LLaMA (3B, 7B), Mistral (Ministral-3B, Mistral-7B), and Gemma (2B, 7B). All methods use 1,000 training examples per identity. These results are reported separately from the primary benchmark in Section~\ref{sec:exp_setup}. For compactness, we report TPR@1\% and AUC in the scaling experiments.

\textit{Identity and family scaling.} 
The nested identity sets start with Qwen-1.5B/3B/7B and LLaMA-3B, then successively add Mistral-3B/7B, LLaMA-7B and Qwen-0.5B, Gemma-2B/7B, and Qwen-14B/32B, yielding 4/6/8/10/12 identities.
With these sets (Table~\ref{tab:expanded_identity_scaling}), \SMark's macro TPR@1\% drops from 0.9986 at 4 identities to a minimum of 0.7462 at 8, then reaches 0.8187 at 12, still above T5 (0.7518) and GPT-2 (0.7188). The advantage is not uniform: T5 leads at 6 identities, and the classifiers attain higher AUC at larger counts.
For family scaling, we select Qwen-3B/7B, LLaMA-3B/7B, Ministral-3B and Mistral-7B, and Gemma-2B/7B as the two identities for each respective family.
Holding these selections fixed and averaging over all family subsets of each size (Table~\ref{tab:expanded_family_scaling}), \SMark leads in TPR@1\% at every family count but trails in AUC. In other words, it is stronger at the calibrated low-FPR operating point than in global ranking.

\begin{table}[!htbp]
\centering
\caption{Identity-count scaling on \textbf{Math} (nested identity sets, 1,000 training examples per identity). Best is bold.}
\label{tab:expanded_identity_scaling}
\small
\setlength{\tabcolsep}{6pt}
\begin{tabular}{lrrrrrr}
\toprule
 & \multicolumn{2}{c}{\textbf{GPT-2}} & \multicolumn{2}{c}{\textbf{T5}} & \multicolumn{2}{c}{\textbf{\textsc{SumMark}}} \\
\cmidrule(lr){2-3}\cmidrule(lr){4-5}\cmidrule(lr){6-7}
\textbf{\#Identities} & \textbf{TPR@1\%} & \textbf{AUC} & \textbf{TPR@1\%} & \textbf{AUC} & \textbf{TPR@1\%} & \textbf{AUC} \\
\midrule
4  & 0.6250 & 0.9735 & 0.8279 & 0.9774 & \textbf{0.9986} & \textbf{0.9987} \\
6  & 0.6213 & 0.9807 & \textbf{0.8089} & \textbf{0.9819} & 0.7811 & 0.9397 \\
8  & 0.5900 & \textbf{0.9792} & 0.7360 & 0.9766 & \textbf{0.7462} & 0.9105 \\
10 & 0.7015 & \textbf{0.9826} & 0.7340 & 0.9765 & \textbf{0.7919} & 0.9339 \\
12 & 0.7188 & \textbf{0.9833} & 0.7518 & 0.9776 & \textbf{0.8187} & 0.9443 \\
\bottomrule
\end{tabular}
\end{table}

\begin{table}[!htbp]
\centering
\caption{Family-count scaling on \textbf{Math}. Each family contributes two identities; results average over all family subsets, so the 4-family point uses 8 identities. Best is bold.}
\label{tab:expanded_family_scaling}
\small
\setlength{\tabcolsep}{6pt}
\begin{tabular}{lrrrrrr}
\toprule
 & \multicolumn{2}{c}{\textbf{GPT-2}} & \multicolumn{2}{c}{\textbf{T5}} & \multicolumn{2}{c}{\textbf{\textsc{SumMark}}} \\
\cmidrule(lr){2-3}\cmidrule(lr){4-5}\cmidrule(lr){6-7}
\textbf{\#Families} & \textbf{TPR@1\%} & \textbf{AUC} & \textbf{TPR@1\%} & \textbf{AUC} & \textbf{TPR@1\%} & \textbf{AUC} \\
\midrule
1 & 0.5788 & \textbf{0.9646} & 0.5825 & 0.9382 & \textbf{0.7850} & 0.9527 \\
2 & 0.6344 & \textbf{0.9769} & 0.6702 & 0.9626 & \textbf{0.8251} & 0.9603 \\
3 & 0.6501 & \textbf{0.9795} & 0.6487 & 0.9670 & \textbf{0.8121} & 0.9596 \\
4 & 0.6977 & \textbf{0.9836} & 0.7061 & 0.9720 & \textbf{0.7935} & 0.9543 \\
\bottomrule
\end{tabular}
\end{table}

\textit{Larger scale gaps.} We test substitutions in which the served model is 4.7--64.0$\times$ smaller than the claimed one (Table~\ref{tab:expanded_gap}). \SMark's macro TPR@1\% is 0.9250 within family and 0.9367 across families, compared with 0.6942/0.8467 for T5 and 0.6467/0.8333 for GPT-2. Detection depends on the specific model pair and does not improve monotonically with the size gap. Within-family substitutions from Qwen-32B achieve TPR@1\% of 0.9900–1.0000, but Qwen-14B$\rightarrow$Qwen-0.5B (0.69) and Qwen-14B$\rightarrow$Ministral-3B (0.74) are weaker, and T5 is perfect on both Ministral-3B directions. These results use public checkpoints and do not establish generalization to proprietary serving stacks.

\begin{table}[!htbp]
\centering
\caption{Identity-Substitution detection with large parameter gaps on \textbf{Math} (1,000 training examples per identity). Red.\ is the claimed-to-served parameter ratio; macro averages are over the six pairs in each panel. Best is bold.}
\label{tab:expanded_gap}
\label{tab:expanded_within_family_gap}
\label{tab:expanded_cross_family_gap}
\small
\setlength{\tabcolsep}{4pt}
\begin{tabular}{lrrrrrrr}
\toprule
 & & \multicolumn{2}{c}{\textbf{GPT-2}} & \multicolumn{2}{c}{\textbf{T5}} & \multicolumn{2}{c}{\textbf{\textsc{SumMark}}} \\
\cmidrule(lr){3-4}\cmidrule(lr){5-6}\cmidrule(lr){7-8}
\textbf{Setting} & \textbf{Red.} & \textbf{TPR@1\%} & \textbf{AUC} & \textbf{TPR@1\%} & \textbf{AUC} & \textbf{TPR@1\%} & \textbf{AUC} \\
\midrule
\multicolumn{8}{l}{\textit{Within family}} \\
Qwen-32B$\rightarrow$Qwen-0.5B & 64.0$\times$ & 0.7800 & 0.9866 & 0.8850 & 0.9924 & \textbf{1.0000} & \textbf{0.9999} \\
Qwen-32B$\rightarrow$Qwen-1.5B & 21.3$\times$ & 0.8250 & 0.9891 & 0.9700 & 0.9986 & \textbf{1.0000} & \textbf{1.0000} \\
Qwen-32B$\rightarrow$Qwen-3B   & 10.7$\times$ & 0.8600 & 0.9912 & 0.9350 & 0.9966 & \textbf{0.9900} & \textbf{0.9997} \\
Qwen-14B$\rightarrow$Qwen-0.5B & 28.0$\times$ & 0.3150 & \textbf{0.9374} & 0.3450 & 0.8949 & \textbf{0.6900} & 0.9278 \\
Qwen-14B$\rightarrow$Qwen-1.5B & 9.3$\times$  & 0.8300 & 0.9926 & \textbf{0.9000} & \textbf{0.9971} & 0.8700 & 0.9667 \\
Qwen-14B$\rightarrow$Qwen-3B   & 4.7$\times$  & 0.2700 & 0.9641 & 0.1300 & 0.9247 & \textbf{1.0000} & \textbf{0.9988} \\
\cmidrule(l){1-8}
Macro average & -- & 0.6467 & 0.9768 & 0.6942 & 0.9674 & \textbf{0.9250} & \textbf{0.9822} \\
\midrule
\multicolumn{8}{l}{\textit{Cross family}} \\
Qwen-32B$\rightarrow$LLaMA-3B     & 10.7$\times$ & 0.8850 & 0.9922 & 0.9900 & 0.9993 & \textbf{1.0000} & \textbf{1.0000} \\
Qwen-32B$\rightarrow$Ministral-3B & 10.7$\times$ & 0.9800 & 0.9952 & \textbf{1.0000} & \textbf{1.0000} & 0.8800 & 0.9745 \\
Qwen-32B$\rightarrow$Gemma-2B     & 16.0$\times$ & 0.9350 & 0.9943 & 0.9800 & 0.9979 & \textbf{1.0000} & \textbf{1.0000} \\
Qwen-14B$\rightarrow$LLaMA-3B     & 4.7$\times$  & 0.3150 & 0.9724 & 0.3050 & 0.9639 & \textbf{1.0000} & \textbf{0.9991} \\
Qwen-14B$\rightarrow$Ministral-3B & 4.7$\times$  & 0.9750 & 0.9985 & \textbf{1.0000} & \textbf{1.0000} & 0.7400 & 0.9532 \\
Qwen-14B$\rightarrow$Gemma-2B     & 7.0$\times$  & 0.9100 & 0.9975 & 0.8050 & 0.9883 & \textbf{1.0000} & \textbf{1.0000} \\
\cmidrule(l){1-8}
Macro average & -- & 0.8333 & \textbf{0.9917} & 0.8467 & 0.9916 & \textbf{0.9367} & 0.9878 \\
\bottomrule
\end{tabular}
\end{table}

\textit{Attribution.} With all 12 identities enrolled, \SMark's exact-identity accuracy is 87.83\% (GPT-2: 82.50\%; T5: 77.38\%; chance: 8.33\%). Mapping the same predictions to families gives 93.17\%; this is not a separate four-class experiment.

\textit{Held-out substitutes.} An auditor cannot enumerate every possible substitute, so we exclude test substitutes from the $L_{\mathrm{sep}}$ negative pool (Table~\ref{tab:heldout_substitutes}). In the held-out-family setting, LLaMA-3B is excluded for the Qwen-3B/7B adapters. In the held-out-scale setting, Qwen-1.5B is excluded for the other adapters. Aggregate TPR@1\% is 0.9990 and 0.9973, respectively. This supports rejecting unseen substitutes under the claimed identity's keyed test, not attributing them. Precision variants are not used as negatives in $L_{\mathrm{sep}}$.

\begin{table}[!htbp]
\centering
\caption{Claim rejection on \textbf{Math} for substitutes held out of the $L_{\mathrm{sep}}$ negative-identity pool. The task is claim rejection, not open-set attribution.}
\label{tab:heldout_substitutes}
\small
\setlength{\tabcolsep}{6pt}
\begin{tabular}{llrr}
\toprule
 & & \multicolumn{2}{c}{\textbf{\textsc{SumMark}}} \\
\cmidrule(lr){3-4}
\textbf{Held-out axis} & \textbf{Setting} & \textbf{TPR@1\%} & \textbf{AUC} \\
\midrule
Family & Qwen-3B$\rightarrow$LLaMA-3B & 0.9990 & 0.9991 \\
       & Qwen-7B$\rightarrow$LLaMA-3B & 0.9990 & 1.0000 \\
       \cmidrule(l){2-4}
       & Aggregate                    & 0.9990 & 0.9995 \\
\midrule
Scale  & Qwen-3B$\rightarrow$Qwen-1.5B  & 1.0000 & 0.9986 \\
       & Qwen-7B$\rightarrow$Qwen-1.5B  & 0.9920 & 0.9984 \\
       & LLaMA-3B$\rightarrow$Qwen-1.5B & 1.0000 & 1.0000 \\
       \cmidrule(l){2-4}
       & Aggregate                      & 0.9973 & 0.9990 \\
\bottomrule
\end{tabular}
\end{table}

\begin{table*}[t]
\centering
\scriptsize
\setlength{\tabcolsep}{2.6pt}
\renewcommand{\arraystretch}{1.10}
\caption{Substitution detection results on \textbf{Math} under different serving precisions. 
For each substitution pair, the provider serves the substituted model at FP16, INT8, or INT4 precision while claiming the target identity.}
\label{tab:precision_robustness}
\resizebox{\textwidth}{!}{%
\begin{tabular}{@{}lccc ccc ccc@{}}
\toprule
& \multicolumn{3}{c}{\textbf{FP16}}
& \multicolumn{3}{c}{\textbf{INT8}}
& \multicolumn{3}{c}{\textbf{INT4}} \\
\cmidrule(lr){2-4}
\cmidrule(lr){5-7}
\cmidrule(lr){8-10}
\textbf{Setting}
& \textbf{TPR@1\%} & \textbf{ACC@1\%} & \textbf{AUC}
& \textbf{TPR@1\%} & \textbf{ACC@1\%} & \textbf{AUC}
& \textbf{TPR@1\%} & \textbf{ACC@1\%} & \textbf{AUC} \\
\midrule

\rowcolor{black!8}
Q7$\rightarrow$Q3
& \textbf{1.0000} & \textbf{0.9950} & \textbf{1.0000}
& \textbf{1.0000} & \textbf{0.9950} & 0.9985
& \textbf{1.0000} & \textbf{0.9950} & 0.9995 \\

\rowcolor{black!8}
Q7$\rightarrow$Q1.5
& 0.9960 & 0.9930 & 0.9974
& \textbf{1.0000} & \textbf{0.9950} & \textbf{0.9995}
& 0.9920 & 0.9910 & 0.9943 \\

\rowcolor{black!8}
Q3$\rightarrow$Q1.5
& \textbf{1.0000} & \textbf{0.9950} & \textbf{0.9998}
& \textbf{1.0000} & \textbf{0.9950} & 0.9985
& 0.9970 & 0.9935 & 0.9945 \\

\rowcolor{black!16}
Q3$\rightarrow$L3
& \textbf{1.0000} & \textbf{0.9950} & \textbf{0.9997}
& \textbf{1.0000} & \textbf{0.9950} & 0.9995
& 0.9970 & 0.9935 & 0.9960 \\

\rowcolor{black!16}
L3$\rightarrow$Q3
& \textbf{1.0000} & \textbf{0.9950} & \textbf{1.0000}
& \textbf{1.0000} & \textbf{0.9950} & 0.9995
& 0.9960 & 0.9930 & 0.9983 \\

\rowcolor{black!22}
Q7$\rightarrow$L3
& \textbf{1.0000} & \textbf{0.9950} & \textbf{1.0000}
& \textbf{1.0000} & \textbf{0.9950} & 0.9995
& 0.9980 & 0.9940 & 0.9994 \\

\rowcolor{black!22}
L3$\rightarrow$Q1.5
& \textbf{1.0000} & \textbf{0.9950} & \textbf{1.0000}
& 0.9970 & 0.9935 & 0.9982
& 0.9860 & 0.9880 & 0.9975 \\

\bottomrule
\end{tabular}
}
\vspace{-0.1in}
\end{table*}

\begin{table*}[t]
\centering

\begin{minipage}[t]{0.49\textwidth}
\vspace{0pt}
\centering
\scriptsize
\setlength{\tabcolsep}{2.0pt}
\renewcommand{\arraystretch}{1.06}
\caption{Ablation on loss components on the Code dataset. \textbf{SFT} uses only $L_{\mathrm{sft}}$, \textbf{+Sep} adds $L_{\mathrm{sep}}$, \textbf{+WM} adds $L_{\mathrm{wm}}$, and \textbf{+Both} adds both losses.}
\label{tab:ablation_loss}
\resizebox{\linewidth}{!}{%
\begin{tabular}{@{}lcccccccc@{}}
\toprule
& \multicolumn{4}{c}{\textbf{TPR@1\%}} 
& \multicolumn{4}{c}{\textbf{AUC}} \\
\cmidrule(lr){2-5}\cmidrule(lr){6-9}
\textbf{Setting} 
& \textbf{SFT} & \textbf{+Sep} & \textbf{+WM} & \textbf{+Both}
& \textbf{SFT} & \textbf{+Sep} & \textbf{+WM} & \textbf{+Both} \\
\midrule

\rowcolor{black!8}
Q7$\rightarrow$Q3
& \textbf{1.0000} & \textbf{1.0000} & \textbf{1.0000} & \textbf{1.0000}
& \textbf{1.0000} & \textbf{1.0000} & \textbf{1.0000} & \textbf{1.0000} \\

\rowcolor{black!8}
Q7$\rightarrow$Q1.5
& 0.9300 & 0.9600 & \textbf{1.0000} & \textbf{1.0000}
& 0.9860 & 0.9930 & \textbf{1.0000} & \textbf{1.0000} \\

\rowcolor{black!8}
Q3$\rightarrow$Q1.5
& \textbf{1.0000} & \textbf{1.0000} & \textbf{1.0000} & \textbf{1.0000}
& \textbf{1.0000} & \textbf{1.0000} & \textbf{1.0000} & \textbf{1.0000} \\

\rowcolor{black!16}
Q3$\rightarrow$L3
& \textbf{1.0000} & \textbf{1.0000} & 0.9800 & \textbf{1.0000}
& \textbf{1.0000} & 0.9999 & 0.9993 & \textbf{1.0000} \\

\rowcolor{black!16}
L3$\rightarrow$Q3
& \textbf{1.0000} & \textbf{1.0000} & 0.9700 & \textbf{1.0000}
& \textbf{1.0000} & 0.9999 & 0.9993 & \textbf{1.0000} \\

\rowcolor{black!22}
Q7$\rightarrow$L3
& \textbf{1.0000} & \textbf{1.0000} & 0.9900 & \textbf{1.0000}
& 0.9999 & \textbf{1.0000} & 0.9953 & \textbf{1.0000} \\

\rowcolor{black!22}
L3$\rightarrow$Q1.5
& 0.9700 & 0.9800 & 0.9600 & \textbf{1.0000}
& 0.9990 & 0.9989 & 0.9992 & \textbf{1.0000} \\

\bottomrule
\end{tabular}
}
\vspace{-0.1in}
\end{minipage}
\hfill
\begin{minipage}[t]{0.49\textwidth}
\vspace{0pt}
\centering
\scriptsize
\setlength{\tabcolsep}{2.0pt}
\renewcommand{\arraystretch}{1.06}
\caption{Comparison of no-watermark (Plain), greenlist watermark (Greenlist), and our method (\SMark) on the \textbf{Math} dataset.}
\label{tab:ablation_wm}
\resizebox{\linewidth}{!}{%
\begin{tabular}{@{}lcccccc@{}}
\toprule
& \multicolumn{3}{c}{\textbf{TPR@1\%}} 
& \multicolumn{3}{c}{\textbf{AUC}} \\
\cmidrule(lr){2-4}\cmidrule(lr){5-7}
\textbf{Setting} 
& \textbf{Plain} & \textbf{Greenlist} & \textbf{\SMark}
& \textbf{Plain} & \textbf{Greenlist} & \textbf{\SMark} \\
\midrule

\rowcolor{black!8}
Q7$\rightarrow$Q3
& 0.0100 & 0.0000 & \textbf{1.0000}
& 0.1717 & 0.1180 & \textbf{1.0000} \\

\rowcolor{black!8}
Q7$\rightarrow$Q1.5
& 0.0000 & 0.0000 & \textbf{0.9960}
& 0.1447 & 0.1080 & \textbf{0.9974} \\

\rowcolor{black!8}
Q3$\rightarrow$Q1.5
& 0.0600 & 0.0310 & \textbf{1.0000}
& 0.5226 & 0.3560 & \textbf{0.9998} \\

\rowcolor{black!16}
Q3$\rightarrow$L3
& 0.0100 & 0.0000 & \textbf{1.0000}
& 0.5052 & 0.1250 & \textbf{0.9997} \\

\rowcolor{black!16}
L3$\rightarrow$Q3
& 0.0400 & 0.0000 & \textbf{1.0000}
& 0.5160 & 0.5170 & \textbf{1.0000} \\

\rowcolor{black!22}
Q7$\rightarrow$L3
& 0.0200 & 0.0000 & \textbf{1.0000}
& 0.3352 & 0.1740 & \textbf{1.0000} \\

\rowcolor{black!22}
L3$\rightarrow$Q1.5
& 0.0400 & 0.0620 & \textbf{1.0000}
& 0.4104 & 0.5890 & \textbf{1.0000} \\

\bottomrule
\end{tabular}
}
\vspace{-0.1in}
\end{minipage}

\vspace{-0.1in}
\end{table*}
\subsection{Robustness Analysis}
\label{app:robustness_full}
\textbf{Lexical watermark removal.} Table~\ref{tab:finetune_removal} evaluates robustness on Code under two downstream perturbations: task-specific summarization fine-tuning and lexical watermark removal. The original \SMark detector achieves near-perfect performance across all substitution pairs. Composing a downstream summarization LoRA with the watermark adapter preserves strong detection in most settings, but reduces TPR@1\% in harder cross-family cases, especially Q3$\rightarrow$L3 and L3$\rightarrow$Q3. This suggests that downstream adaptation can shift the summary token distribution and narrow the low-FPR detection margin. In contrast, the copy-paste watermark removal attack has little effect: TPR@1\% remains between 0.997 and 1.000 with AUC ranging from 0.9973 to 0.9995. This indicates that \SMark’s signal is distributed across summary-level token statistics rather than concentrated in easily removable local spans.

\begin{table*}[t]
\centering
\small
\setlength{\tabcolsep}{3.2pt}
\renewcommand{\arraystretch}{1.12}
\caption{Robustness on Code under downstream fine-tuning and lexical watermark removal. For fine-tuning, we compose a task-specific summarization LoRA with the watermark adapter at inference. For watermark removal, we apply a copy-paste attack that replaces contiguous spans in the summary.}
\label{tab:finetune_removal}
\resizebox{\textwidth}{!}{%
\begin{tabular}{@{}lccc ccc ccc@{}}
\toprule
& \multicolumn{3}{c}{\textbf{Original}}
& \multicolumn{3}{c}{\textbf{Model Finetuning}}
& \multicolumn{3}{c}{\textbf{Watermark Removal}} \\
\cmidrule(lr){2-4}\cmidrule(lr){5-7}\cmidrule(lr){8-10}
\textbf{Setting}
& \textbf{TPR@1\%} & \textbf{ACC@1\%} & \textbf{AUC}
& \textbf{TPR@1\%} & \textbf{ACC@1\%} & \textbf{AUC}
& \textbf{TPR@1\%} & \textbf{ACC@1\%} & \textbf{AUC} \\
\midrule

\rowcolor{black!8}
Q7$\rightarrow$Q3
& \textbf{1.0000} & \textbf{0.9950} & \textbf{1.0000}
& 0.9630 & 0.9765 & 0.9950
& \textbf{1.0000} & \textbf{0.9950} & 0.9995 \\

\rowcolor{black!8}
Q7$\rightarrow$Q1.5
& \textbf{1.0000} & \textbf{0.9950} & \textbf{1.0000}
& 0.9770 & 0.9835 & 0.9978
& 0.9970 & 0.9935 & 0.9973 \\

\rowcolor{black!8}
Q3$\rightarrow$Q1.5
& \textbf{1.0000} & \textbf{0.9950} & \textbf{1.0000}
& 0.9780 & 0.9840 & 0.9975
& \textbf{1.0000} & \textbf{0.9950} & 0.9994 \\

\rowcolor{black!16}
Q3$\rightarrow$L3
& \textbf{1.0000} & \textbf{0.9950} & \textbf{1.0000}
& 0.4480 & 0.7190 & 0.9553
& \textbf{1.0000} & \textbf{0.9950} & 0.9994 \\

\rowcolor{black!16}
L3$\rightarrow$Q3
& \textbf{1.0000} & \textbf{0.9950} & \textbf{1.0000}
& 0.7160 & 0.8530 & 0.9832
& \textbf{1.0000} & \textbf{0.9950} & 0.9990 \\

\rowcolor{black!22}
Q7$\rightarrow$L3
& \textbf{1.0000} & \textbf{0.9950} & \textbf{1.0000}
& 0.9840 & 0.9870 & 0.9979
& \textbf{1.0000} & \textbf{0.9950} & 0.9995 \\

\rowcolor{black!22}
L3$\rightarrow$Q1.5
& \textbf{1.0000} & \textbf{0.9950} & \textbf{1.0000}
& 0.9640 & 0.9770 & 0.9956
& 0.9990 & 0.9945 & 0.9990 \\

\bottomrule
\end{tabular}
}
\vspace{-0.1in}
\end{table*}

\begin{table}[t]
\centering
\small
\caption{Robustness under spoofing attacks on Math. We evaluate two text-only spoofing strategies: distilled adapter spoofing and watermark-stealing spoofing.
For both attacks, lower ASR@1\% and higher TPR@1\%/AUC indicate stronger audit robustness.}
\label{tab:spoofing}
\resizebox{0.8\columnwidth}{!}{%
\begin{tabular}{@{}lccc ccc@{}}
\toprule
& \multicolumn{3}{c}{\textbf{Distilled Adapter Spoofing}} 
& \multicolumn{3}{c}{\textbf{Watermark Stealing Spoofing}} \\
\cmidrule(lr){2-4}\cmidrule(lr){5-7}
\textbf{Setting}
& \textbf{TPR@1\%} & \textbf{ASR@1\%} & \textbf{AUC}
& \textbf{TPR@1\%} & \textbf{ASR@1\%} & \textbf{AUC} \\
\midrule

\rowcolor{black!8}
Q7$\rightarrow$Q3
& 0.9980 & 0.0020 & 0.9999
& 0.7840 & 0.2160 & 0.9930 \\

\rowcolor{black!8}
Q7$\rightarrow$Q1.5
& 0.9830 & 0.0170 & 0.9939
& 0.9240 & 0.0760 & 0.9960 \\

\rowcolor{black!8}
Q3$\rightarrow$Q1.5
& 0.9960 & 0.0040 & 0.9983
& 1.0000 & 0.0000 & 1.0000 \\

\rowcolor{black!16}
Q3$\rightarrow$L3
& 0.9570 & 0.0430 & 0.9938
& 1.0000 & 0.0000 & 1.0000 \\

\rowcolor{black!16}
L3$\rightarrow$Q3
& 0.3710 & 0.6290 & 0.9474
& 1.0000 & 0.0000 & 1.0000 \\

\rowcolor{black!22}
Q7$\rightarrow$L3
& 0.7500 & 0.2500 & 0.9706
& 1.0000 & 0.0000 & 1.0000 \\

\rowcolor{black!22}
L3$\rightarrow$Q1.5
& 0.7890 & 0.2110 & 0.9641
& 1.0000 & 0.0000 & 1.0000 \\

\bottomrule
\end{tabular}
}
\vspace{-0.1in}
\end{table}

\textbf{Watermark spoofing.}
Table~\ref{tab:spoofing} evaluates two text-only spoofing attacks where the adversary observes watermarked summaries from the claimed identity but has no access to the auditor's secret key, carrier sets, or adapter weights.
In \emph{distilled adapter spoofing}, the adversary collects watermarked prompt--summary pairs from the claimed identity and trains a local spoof adapter on the substituted model to imitate the claimed identity's summary distribution. In \emph{watermark-stealing spoofing}, the adversary estimates token-level carrier statistics from observed watermarked summaries and biases decoding toward tokens that appear likely to belong to the claimed identity's carrier set.

Distilled adapter spoofing is mostly contained: ASR@1\% stays below 0.05 for four of seven transfer directions, with TPR@1\% between 0.9570 and 0.9980. This suggests that spoofing requires approximating the claimed identity's keyed statistical footprint, not merely its summary style. However, some directions are substantially harder, especially L3$\rightarrow$Q3 (ASR@1\% = 0.6290), Q7$\rightarrow$L3 (0.2500), and L3$\rightarrow$Q1.5 (0.2110).

Watermark-stealing spoofing is less effective: ASR@1\% is zero in most settings, except Q7$\rightarrow$Q3 (0.2160) and Q7$\rightarrow$Q1.5 (0.0760). Overall, \SMark is robust to text-only watermark stealing and to distilled spoofing in most directions, but adaptive distilled spoofing remains a limitation in specific transfer directions.

\textbf{Summary length sensitivity.} Table~\ref{tab:length_sensitivity} evaluates how detection changes when released summaries are truncated to different prefix lengths. Overall, \SMark remains effective even with short summaries: at 128 tokens, six of seven settings already achieve TPR@1\% at least 0.925, and several settings remain perfect across all lengths. This suggests that the identity footprint is not dependent on unusually long summaries.

The main exception is Q3$\rightarrow$L3, where TPR@1\% stays around 0.65--0.70 despite high AUC. This indicates that the two distributions remain separable in aggregate, but the margin is not large enough to support near-perfect detection at the strict 1\% FPR threshold. Performance is also not strictly monotonic with length, since longer prefixes may introduce additional natural lexical variation rather than only more watermark evidence. Overall, the results show that summary length affects the low-FPR margin, but short summaries are sufficient for reliable detection in most substitution directions.
\begin{table*}[t]
\centering
\scriptsize
\setlength{\tabcolsep}{2.4pt}
\renewcommand{\arraystretch}{1.10}
\caption{Length sensitivity of substitution detection on Code. We truncate each released summary to the first 128, 256, 384, or 512 tokens, and report TPR@FPR=1\%, ACC@FPR=1\%, and AUC. Setting A$\rightarrow$B denotes serving B while claiming A.}
\label{tab:length_sensitivity}
\resizebox{\textwidth}{!}{%
\begin{tabular}{@{}lccc ccc ccc ccc@{}}
\toprule
& \multicolumn{3}{c}{\textbf{128 tokens}}
& \multicolumn{3}{c}{\textbf{256 tokens}}
& \multicolumn{3}{c}{\textbf{384 tokens}}
& \multicolumn{3}{c}{\textbf{512 tokens}} \\
\cmidrule(lr){2-4}
\cmidrule(lr){5-7}
\cmidrule(lr){8-10}
\cmidrule(lr){11-13}
\textbf{Setting}
& \textbf{TPR@1\%} & \textbf{ACC@1\%} & \textbf{AUC}
& \textbf{TPR@1\%} & \textbf{ACC@1\%} & \textbf{AUC}
& \textbf{TPR@1\%} & \textbf{ACC@1\%} & \textbf{AUC}
& \textbf{TPR@1\%} & \textbf{ACC@1\%} & \textbf{AUC} \\
\midrule

\rowcolor{black!8}
Q7$\rightarrow$Q3
& \textbf{1.0000} & \textbf{1.0000} & \textbf{1.0000}
& \textbf{1.0000} & \textbf{1.0000} & \textbf{1.0000}
& \textbf{1.0000} & \textbf{1.0000} & \textbf{1.0000}
& \textbf{1.0000} & \textbf{1.0000} & \textbf{1.0000} \\

\rowcolor{black!8}
Q7$\rightarrow$Q1.5
& 0.9250 & 0.9625 & 0.9944
& \textbf{1.0000} & \textbf{1.0000} & \textbf{1.0000}
& 0.9750 & 0.9875 & 0.9969
& 0.9750 & 0.9875 & 0.9994 \\

\rowcolor{black!8}
Q3$\rightarrow$Q1.5
& \textbf{1.0000} & \textbf{1.0000} & \textbf{1.0000}
& \textbf{1.0000} & \textbf{1.0000} & \textbf{1.0000}
& 0.7750 & 0.8875 & 0.9944
& 0.9750 & 0.9875 & 0.9981 \\

\rowcolor{black!16}
Q3$\rightarrow$L3
& \textbf{0.7000} & \textbf{0.8500} & 0.9825
& \textbf{0.7000} & \textbf{0.8500} & 0.9788
& 0.6750 & 0.8375 & \textbf{0.9856}
& 0.6500 & 0.8250 & 0.9756 \\

\rowcolor{black!16}
L3$\rightarrow$Q3
& 0.9750 & 0.9875 & 0.9938
& 0.9750 & 0.9875 & 0.9925
& \textbf{1.0000} & \textbf{1.0000} & \textbf{1.0000}
& 0.9750 & 0.9875 & 0.9994 \\

\rowcolor{black!22}
Q7$\rightarrow$L3
& \textbf{1.0000} & \textbf{1.0000} & \textbf{1.0000}
& \textbf{1.0000} & \textbf{1.0000} & \textbf{1.0000}
& \textbf{1.0000} & \textbf{1.0000} & \textbf{1.0000}
& \textbf{1.0000} & \textbf{1.0000} & \textbf{1.0000} \\

\rowcolor{black!22}
L3$\rightarrow$Q1.5
& \textbf{0.9750} & \textbf{0.9875} & \textbf{0.9963}
& \textbf{0.9750} & \textbf{0.9875} & 0.9931
& 0.8500 & 0.9250 & 0.9644
& 0.9000 & 0.9500 & 0.9656 \\

\bottomrule
\end{tabular}
}
\vspace{-0.1in}
\end{table*}

\begin{table}
\caption{
Robustness of watermark detection under summary post-processing and rewriting on Math.
We report three bounded surface transformations
(sentence shuffle, random deletion, and word truncation), and one stronger bilingual rewriting stress test.
}
\label{tab:rewriting}
\resizebox{\textwidth}{!}{%
\begin{tabular}{@{}lcccccccccc@{}}
\toprule
& \multicolumn{2}{c}{\textbf{Original}}
& \multicolumn{2}{c}{\textbf{Sentence Shuffle}}
& \multicolumn{2}{c}{\textbf{Random Deletion}}
& \multicolumn{2}{c}{\textbf{Truncate Words}}
& \multicolumn{2}{c}{\textbf{Bilingual Rewrite}} \\
\cmidrule(lr){2-3}
\cmidrule(lr){4-5}
\cmidrule(lr){6-7}
\cmidrule(lr){8-9}
\cmidrule(lr){10-11}
\textbf{Setting}
& \textbf{TPR@1\%} & \textbf{AUC}
& \textbf{TPR@1\%} & \textbf{AUC}
& \textbf{TPR@1\%} & \textbf{AUC}
& \textbf{TPR@1\%} & \textbf{AUC}
& \textbf{TPR@1\%} & \textbf{AUC} \\
\midrule

\rowcolor{black!8}
Q7$\rightarrow$Q3
& \textbf{1.0000} & \textbf{1.0000}
& \textbf{1.0000} & 0.9995
& \textbf{1.0000} & 0.9995
& \textbf{1.0000} & 0.9995
& 0.9760 & 0.9963 \\

\rowcolor{black!8}
Q7$\rightarrow$Q1.5
& \textbf{0.9960} & 0.9974
& \textbf{0.9960} & 0.9969
& 0.9950 & 0.9969
& 0.9950 & \textbf{0.9981}
& 0.9010 & 0.9931 \\

\rowcolor{black!8}
Q3$\rightarrow$Q1.5
& \textbf{1.0000} & \textbf{0.9998}
& \textbf{1.0000} & 0.9994
& \textbf{1.0000} & 0.9993
& \textbf{1.0000} & 0.9994
& 0.8220 & 0.9840 \\

\rowcolor{black!16}
Q3$\rightarrow$L3
& \textbf{1.0000} & \textbf{0.9997}
& \textbf{1.0000} & 0.9993
& \textbf{1.0000} & 0.9992
& \textbf{1.0000} & 0.9994
& 0.6570 & 0.9652 \\

\rowcolor{black!16}
L3$\rightarrow$Q3
& \textbf{1.0000} & \textbf{1.0000}
& \textbf{1.0000} & 0.9990
& \textbf{1.0000} & 0.9990
& \textbf{1.0000} & 0.9990
& 0.6390 & 0.9646 \\

\rowcolor{black!22}
Q7$\rightarrow$L3
& \textbf{1.0000} & \textbf{1.0000}
& \textbf{1.0000} & 0.9995
& \textbf{1.0000} & 0.9995
& \textbf{1.0000} & 0.9995
& 0.9610 & 0.9961 \\

\rowcolor{black!22}
L3$\rightarrow$Q1.5
& \textbf{1.0000} & \textbf{1.0000}
& \textbf{1.0000} & 0.9990
& 0.9990 & 0.9989
& 0.9990 & 0.9989
& 0.5740 & 0.9388 \\

\bottomrule
\end{tabular}
}
\vspace{-0.1in}
\end{table}

\textbf{Summary post-processing.}
Table~\ref{tab:rewriting} reports per-pair detection results on Math under sentence shuffling, random deletion, word truncation, and bilingual rewriting. Across the three bounded surface transformations, the average TPR@1\% over all seven substitution pairs and three transformations is 0.9992, with a minimum of 0.9950. Bilingual rewriting reduces the average TPR@1\% to 0.7900 and the minimum to 0.5740. These results indicate that detection remains stable under the evaluated bounded edits, while more extensive rewriting can weaken the watermark signal.

\textbf{Cross-domain generalization.}
We train each adapter on Math prompts only, freeze it together with the Math-calibrated thresholds, and evaluate the same seven substitution pairs on Code and General without retraining or recalibration (Table~\ref{tab:domain_transfer}). Macro TPR remains above 0.998 on both target domains, with observed macro FPR of 0.0061 on Code and 0.0123 on General. Per-pair FPR ranges from 0.001 to 0.014 on Code and from 0.002 to 0.025 on General, so calibration drift is pair-dependent and some General directions exceed the nominal 1\% target. These results cover two task-domain shifts only, not arbitrary shifts or an adaptive provider that alters the generation process.

\begin{table}[!htbp]
\centering
\caption{Cross-domain transfer from \textbf{Math} to \textbf{Code} and \textbf{General}. Adapters and detection thresholds are trained and calibrated on Math and kept fixed; observed FPR is measured on target-domain data. Metrics are macro-averaged over seven substitution pairs.}
\label{tab:domain_transfer}
\small
\setlength{\tabcolsep}{6pt}
\begin{tabular}{lrrrr}
\toprule
 & \multicolumn{4}{c}{\textbf{\textsc{SumMark}}} \\
\cmidrule(lr){2-5}
\textbf{Train$\rightarrow$test} & \textbf{TPR} & \textbf{Observed FPR} & \textbf{Balanced ACC} & \textbf{AUC} \\
\midrule
Math$\rightarrow$Math    & 0.9994 & 0.0100 & 0.9947 & 0.9996 \\
Math$\rightarrow$Code    & 0.9993 & 0.0061 & 0.9966 & 0.9991 \\
Math$\rightarrow$General & 0.9989 & 0.0123 & 0.9933 & 0.9990 \\
\bottomrule
\end{tabular}
\end{table}

\textbf{Training data poisoning.}
Trusted onboarding assumes an uncorrupted adapter-training corpus. We stress-test this assumption on Qwen-7B$\rightarrow$Qwen-3B by flipping the carrier identity label between the two models for a fraction of the 1,000 training examples. Prompts and teacher summaries are unchanged, so the $L_{\mathrm{wm}}$ and $L_{\mathrm{sep}}$ targets are corrupted while $L_{\mathrm{sft}}$ is not (Table~\ref{tab:training_poisoning}). Detection is nearly unaffected at 5\% poisoning, degrades at 15\%--25\%, and TPR@1\% drops to zero from 50\% onward. At 50\%, AUC is still 0.8767: some score ranking survives, but detection at the calibrated low-FPR threshold does not. This marks a training-integrity boundary rather than a defense against poisoning; authenticated data provenance lies outside the verifier studied here.

\begin{table}[!htbp]
\centering
\caption{Carrier-label poisoning on Qwen-7B$\rightarrow$Qwen-3B (\textbf{Math}, 1,000 training examples per rate). Prompts and teacher summaries are unchanged; only the $L_{\mathrm{wm}}$ and $L_{\mathrm{sep}}$ carrier targets are corrupted. At 100\%, labels are fully swapped and the score ordering is inverted. }
\label{tab:training_poisoning}
\small
\setlength{\tabcolsep}{6pt}
\begin{tabular}{lrrr}
\toprule
 & \multicolumn{3}{c}{\textbf{\textsc{SumMark}}} \\
\cmidrule(lr){2-4}
\textbf{Poison rate} & \textbf{TPR@1\%} & \textbf{ACC@1\%} & \textbf{AUC} \\
\midrule
0\%   & 1.0000 & 0.9950 & 1.0000 \\
5\%   & 0.9900 & 0.9925 & 0.9998 \\
15\%  & 0.7950 & 0.8925 & 0.9761 \\
25\%  & 0.6900 & 0.8400 & 0.9602 \\
50\%  & 0.0000 & 0.4950 & 0.8767 \\
100\% & 0.0000 & 0.4950 & 0.0032 \\
\bottomrule
\end{tabular}
\end{table}

\subsection{Ablation Studies}
\label{app:ablations}

\textbf{Ablation 1: Do we need identity-specific watermarking?}
Table~\ref{tab:ablation_wm} compares three variants on Math:\textbf{Plain} (no watermark), a \textbf{Greenlist} watermark that uses a \emph{single shared} greenlist across identities, and \textbf{\SMark}. Under a strict auditing regime (TPR at FPR$=1\%$), both Plain and Greenlist fail across all substitution settings, even when AUC appears moderate. This highlights a core requirement of model \emph{identity} auditing: a global greenlist can create a detectable bias in aggregate, but it does not provide \emph{separable, identity-specific} evidence needed to reliably distinguish a claimed identity from closely related alternatives at low false-positive rates.

\textit{Why Greenlist fails at low FPR.} Math summaries often have constrained lexical choices driven by the solution structure, leaving limited room for a weak global token bias to accumulate. As a result, non-watermarked summaries can match a shared greenlist by chance, and the separation becomes insufficient when the threshold is set to control false alarms. By contrast, \SMark explicitly allocates \emph{identity-specific} carrier sets and trains adapters to concentrate additional probability mass on the correct bucket, which yields a predictable shift in summary token statistics that remains separable across identities.

\textbf{Ablation 2: Training objective components.} We next isolate the role of the detectability and separability components in our training objective by ablating $L_{\mathrm{wm}}$ (green-mass regularization) and $L_{\mathrm{sep}}$ (margin-based separation across identities). As shown in Table~\ref{tab:ablation_loss} and Figure~\ref{fig:para_loss_1} when varying weights, removing either term can reduce TPR at FPR$=1\%$, even when AUC remains high. This reflects the difference between \emph{overall separability} and \emph{auditing at strict operating points}. Concretely, $L_{\mathrm{wm}}$ supplies the primary watermark signal by increasing usage of the correct identity bucket, while $L_{\mathrm{sep}}$ suppresses the most confusable competing identity and stabilizes decisions across substitution pairs. Combining both terms achieves consistently strong low-FPR detection, matching the design goal of robust identity attribution under closely related substitutions.

\textbf{Ablation 3: Effect of precision-consistency regularizer.}
\label{app:ablation_lqs}
Table~\ref{tab:ablation_lqs} shows that the precision-consistency regularizer $L_{\mathrm{qs}}$ is essential for detecting precision variants. Without $L_{\mathrm{qs}}$, the detector is almost random: TPR@1\% is near zero, ACC@1\% stays around 0.50, and AUC is close to chance across all models. After adding $L_{\mathrm{qs}}$, detection improves sharply for most identities. LLaMA-3B, Qwen-3B, and Qwen-7B all achieve near-perfect TPR@1\% and AUC for both FP16$\rightarrow$INT8 and FP16$\rightarrow$INT4.

The main exception is Qwen-1.5B, where INT8 remains difficult and INT4 is only partially detectable. Still, $L_{\mathrm{qs}}$ substantially improves its AUC, especially for FP16$\rightarrow$INT4. Overall, these results show that standard identity watermarking alone does not naturally separate FP16 and quantized variants; the quantization-separation loss is what creates a precision-sensitive audit signal.

\textbf{Ablation 4: Adapter-matched text and watermarking baselines.}
The shared-Greenlist and generic watermark baselines above do not control for the use of an identity-specific adapter. We therefore add adapter-matched baselines on Math, using the same seven substitution pairs and 1,000 training examples per adapter. \emph{T5 + SFT adapter} trains an SFT-only summary adapter on clean, unwatermarked reference summaries; T5 is then trained and evaluated on its outputs alone, without a key or carrier information. \emph{Keyed Unigram/Greenlist adapters} assign each identity its own secret key and train an adapter to reproduce teacher summaries watermarked with that key. Both controls use an adapter per identity, and the keyed variants a separate key per identity, but neither uses \textsc{SumMark}'s controlled-overlap carriers or hardest-negative separability objective.

As shown in Table~\ref{tab:adapter_matched_controls}, the strongest control, T5 + SFT adapter, reaches a mean TPR@1\% of 0.8937 but drops to 0.6550 on its worst pair. The keyed watermark adapters perform far worse, while \textsc{SumMark} reaches 0.9980. Under these implementations and budget, the gain is therefore not explained by adapter-based generation or per-identity keys alone, although this does not rule out multi-key watermark attribution in general. This control differs from the SFT-only loss ablation in Table~\ref{tab:ablation_loss}, which removes \textsc{SumMark}'s own training terms.

\begin{table}[!htbp]
\centering
\caption{Adapter-matched baselines on \textbf{Math} (1,000 training examples per identity). Mean and minimum are taken over seven substitution pairs. T5 + SFT adapter uses unwatermarked summaries; keyed Unigram/Greenlist adapters use a separate key and adapter per identity. Best is bold.}
\label{tab:adapter_matched_controls}
\small
\setlength{\tabcolsep}{6pt}
\begin{tabular}{lrrrr}
\toprule
 & \multicolumn{2}{c}{\textbf{TPR@1\%}} & \multicolumn{2}{c}{\textbf{AUC}} \\
\cmidrule(lr){2-3}\cmidrule(lr){4-5}
\textbf{Method} & \textbf{Mean} & \textbf{Min} & \textbf{Mean} & \textbf{Min} \\
\midrule
T5 + SFT adapter        & 0.8937 & 0.6550 & 0.9899 & 0.9619 \\
Keyed Unigram adapter   & 0.3328 & 0.0020 & 0.7890 & 0.1340 \\
Keyed Greenlist adapter & 0.0163 & 0.0000 & 0.5319 & 0.4823 \\
\textsc{SumMark}        & \textbf{0.9980} & \textbf{0.9920} & \textbf{0.9981} & \textbf{0.9970} \\
\bottomrule
\end{tabular}
\end{table}

\begin{figure}[t]
    \centering
    \includegraphics[width=0.6\linewidth]{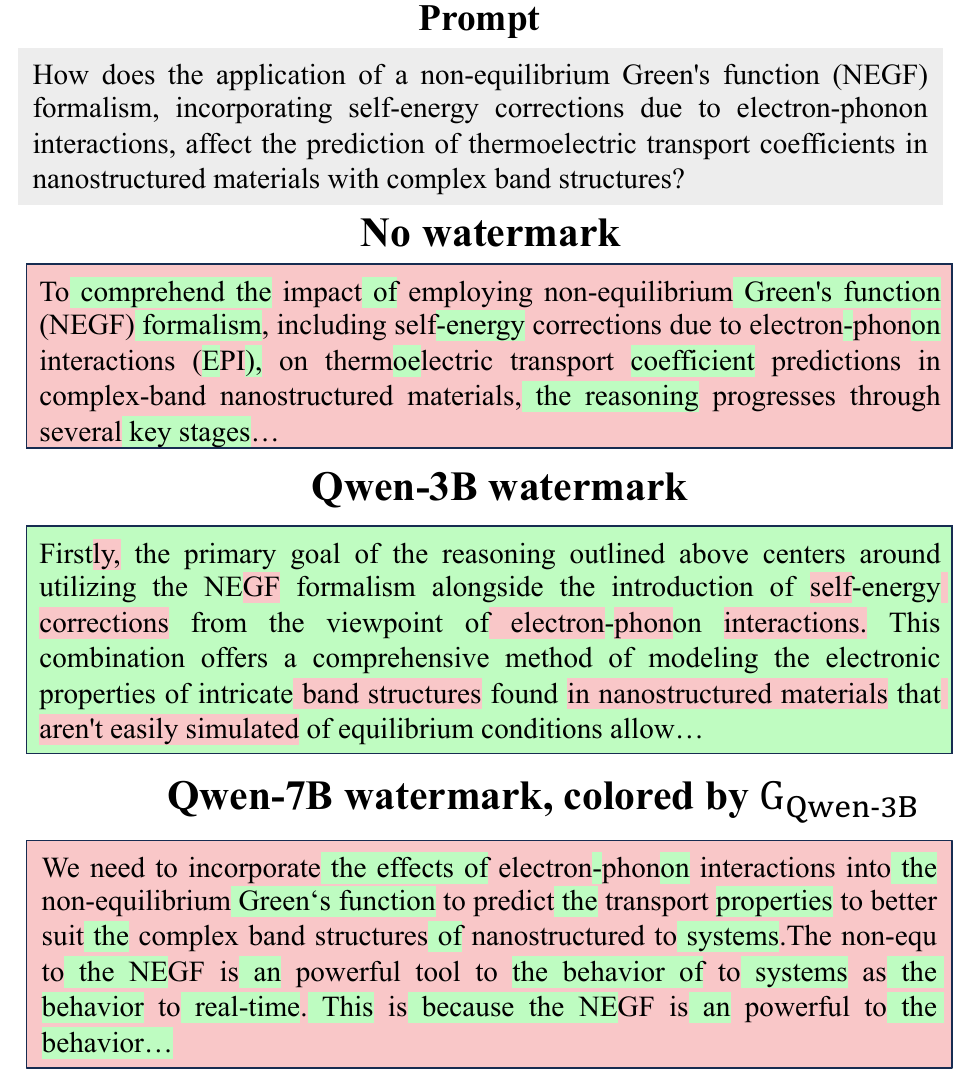}
    \vspace{-0.1in}
    \caption{Visualizing identity footprints: tokens in $G_{\text{Qwen-3B}}$ are highlighted. The Qwen-3B adapter yields a denser, more uniform green-token pattern than no-watermark, while a substituted identity (Qwen-7B) does not reproduce this footprint when colored by $G_{\text{Qwen-3B}}$.}
    \label{fig:case}
    \vspace{-0.15in}
\end{figure}

\begin{table}[t]
\centering
\small
\setlength{\tabcolsep}{3pt}
\renewcommand{\arraystretch}{1.12}
\caption{Effect of the quantization-separation loss $L_{\mathrm{qs}}$ on within-identity precision-consistency detection.
We compare watermark adapters trained with and without $L_{\mathrm{qs}}$ when the claimed configuration is FP16 and the served configuration is INT8 or INT4.}
\label{tab:ablation_lqs}
\resizebox{0.85\columnwidth}{!}{%
\begin{tabular}{@{}llccc ccc@{}}
\toprule
\multirow{2}{*}{\textbf{Model}} 
& \multirow{2}{*}{\textbf{Setting}} 
& \multicolumn{3}{c}{\textbf{w/o $L_{\mathrm{qs}}$}}
& \multicolumn{3}{c}{\textbf{w/ $L_{\mathrm{qs}}$}} \\
\cmidrule(lr){3-5}\cmidrule(lr){6-8}
& 
& \textbf{TPR@1\%} & \textbf{ACC@1\%} & \textbf{AUC}
& \textbf{TPR@1\%} & \textbf{ACC@1\%} & \textbf{AUC} \\
\midrule

\multirow{2}{*}{LLaMA-3B}
& FP16$\rightarrow$INT8
& 0.0150 & 0.5040 & 0.5034
& \textbf{0.9990} & \textbf{0.9960} & \textbf{0.9990} \\
& FP16$\rightarrow$INT4
& 0.0000 & 0.5000 & 0.4256
& \textbf{0.9990} & \textbf{0.9995} & \textbf{0.9990} \\

\midrule
\multirow{2}{*}{Qwen-1.5B}
& FP16$\rightarrow$INT8
& 0.0090 & 0.5000 & 0.4952
& \textbf{0.1132} & \textbf{0.5521} & \textbf{0.7575} \\
& FP16$\rightarrow$INT4
& 0.0181 & 0.5050 & 0.4875
& \textbf{0.3417} & \textbf{0.6663} & \textbf{0.9209} \\

\midrule
\multirow{2}{*}{Qwen-3B}
& FP16$\rightarrow$INT8
& 0.0020 & 0.5000 & 0.5298
& \textbf{0.9860} & \textbf{0.9890} & \textbf{0.9973} \\
& FP16$\rightarrow$INT4
& 0.0000 & 0.5000 & 0.3719
& \textbf{0.9960} & \textbf{0.9945} & \textbf{0.9986} \\

\midrule
\multirow{2}{*}{Qwen-7B}
& FP16$\rightarrow$INT8
& 0.0140 & 0.5030 & 0.4868
& \textbf{0.9910} & \textbf{0.9920} & \textbf{0.9982} \\
& FP16$\rightarrow$INT4
& 0.0140 & 0.5045 & 0.5611
& \textbf{0.9990} & \textbf{0.9975} & \textbf{0.9998} \\

\bottomrule
\end{tabular}
}
\vspace{-0.1in}
\end{table}

\subsection{Case Study}
\label{sec:case_study}
To complement the quantitative ablations and robustness tests above, we provide a qualitative visualization of what the identity watermark looks like in a single summary. Figure~\ref{fig:case} highlights tokens from a fixed bucket to show how the correct adapter produces a consistently denser, more uniformly distributed footprint, while a substituted identity does not. This case study visualizes identity-specific watermarking at the token level under three settings: no watermark, Qwen-3B watermark, and Qwen-7B watermark. For visualization, all texts are colored using the same vocabulary bucket $G_{\text{Qwen-3B}}$, where tokens in this set are marked green. Without watermarking, green tokens appear only occasionally, consistent with natural generation. With the Qwen-3B watermark enabled, green tokens become noticeably more frequent and more evenly spread throughout the summary, indicating a stable, detectable bias. We color Qwen-7B summaries by $G_{\text{Qwen-3B}}$ to simulate a substitution case (serving Qwen-7B while claiming Qwen-3B); the resulting summary does not exhibit the same green-token pattern, illustrating that watermark footprints are separable across model identities.

\subsection{Hyperparameter Analysis}
\label{app:hyper}

\textbf{Substitution ratio.}
To ensure fair comparison, we fix the total number of test samples to 1,000 for each substitution ratio across all three datasets. As shown in Tables~\ref{tab:sub_ratio_math}, and ~\ref{tab:sub_ratio_general}, detection performance consistently improves as the substitution ratio increases on Math and General. Even at low substitution ratios, TPR@FPR=1\% is close to 1.0 for most substitution settings, and AUC remains around 0.99. At a substitution ratio of 0.5, performance is effectively saturated, with near-perfect TPR and consistently high AUC across all model pairs. We therefore adopt a ratio of 0.5 as the default setting in our experiments.

\textbf{}

\begin{table*}[t]
\centering
\small
\caption{Sensitivity to \textbf{Substitution Ratio} on \textbf{Math}.
We report TPR@FPR=1\%, ACC@FPR=1\%, and AUC.
Q7/Q3/Q1.5 = Qwen-7B/3B/1.5B, L3 = LLaMA-3B.
Best is bold.}
\setlength{\tabcolsep}{2.2pt}
\renewcommand{\arraystretch}{1.06}
\resizebox{\linewidth}{!}{%
\begin{tabular}{l ccc ccc ccc ccc ccc ccc}
\toprule
& \multicolumn{3}{c}{\textbf{5\%}}
& \multicolumn{3}{c}{\textbf{10\%}}
& \multicolumn{3}{c}{\textbf{15\%}}
& \multicolumn{3}{c}{\textbf{25\%}}
& \multicolumn{3}{c}{\textbf{50\%}}
& \multicolumn{3}{c}{\textbf{65\%}} \\
\cmidrule(lr){2-4}\cmidrule(lr){5-7}\cmidrule(lr){8-10}\cmidrule(lr){11-13}\cmidrule(lr){14-16}\cmidrule(lr){17-19}
\textbf{Setting}
& \textbf{TPR} & \textbf{ACC} & \textbf{AUC}
& \textbf{TPR} & \textbf{ACC} & \textbf{AUC}
& \textbf{TPR} & \textbf{ACC} & \textbf{AUC}
& \textbf{TPR} & \textbf{ACC} & \textbf{AUC}
& \textbf{TPR} & \textbf{ACC} & \textbf{AUC}
& \textbf{TPR} & \textbf{ACC} & \textbf{AUC} \\
\midrule

\rowcolor{black!8}
Q7$\to$Q3
& \textbf{1.0000} & 0.9910 & \textbf{0.9995}
& \textbf{1.0000} & 0.9910 & 0.9994
& \textbf{1.0000} & 0.9920 & 0.9994
& \textbf{1.0000} & 0.9930 & 0.9993
& \textbf{1.0000} & 0.9950 & 0.9990
& \textbf{1.0000} & \textbf{0.9970} & 0.9986 \\

\rowcolor{black!8}
Q7$\to$Q1.5
& \textbf{1.0000} & 0.9910 & \textbf{0.9995}
& \textbf{1.0000} & 0.9910 & 0.9994
& 0.9933 & 0.9910 & 0.9931
& 0.9920 & 0.9910 & 0.9955
& 0.9940 & 0.9920 & 0.9953
& 0.9969 & \textbf{0.9950} & 0.9957 \\

\rowcolor{black!8}
Q3$\to$Q1.5
& \textbf{1.0000} & 0.9910 & \textbf{0.9994}
& \textbf{1.0000} & 0.9910 & 0.9989
& \textbf{1.0000} & 0.9920 & 0.9990
& \textbf{1.0000} & 0.9930 & 0.9985
& \textbf{1.0000} & 0.9950 & 0.9989
& \textbf{1.0000} & \textbf{0.9970} & 0.9985 \\

\rowcolor{black!16}
Q3$\to$L3
& \textbf{1.0000} & 0.9910 & \textbf{0.9994}
& \textbf{1.0000} & 0.9910 & 0.9992
& \textbf{1.0000} & 0.9920 & 0.9993
& \textbf{1.0000} & 0.9930 & 0.9992
& \textbf{1.0000} & 0.9950 & 0.9988
& \textbf{1.0000} & \textbf{0.9970} & 0.9940 \\

\rowcolor{black!16}
L3$\to$Q3
& \textbf{1.0000} & 0.9910 & \textbf{0.9990}
& \textbf{1.0000} & 0.9910 & 0.9989
& \textbf{1.0000} & 0.9920 & 0.9988
& \textbf{1.0000} & 0.9930 & 0.9987
& \textbf{1.0000} & 0.9950 & \textbf{0.9990}
& \textbf{1.0000} & \textbf{0.9970} & 0.9986 \\

\rowcolor{black!22}
Q7$\to$L3
& \textbf{1.0000} & 0.9910 & \textbf{0.9995}
& \textbf{1.0000} & 0.9910 & \textbf{0.9995}
& \textbf{1.0000} & 0.9920 & 0.9994
& \textbf{1.0000} & 0.9930 & 0.9993
& \textbf{1.0000} & 0.9950 & 0.9990
& \textbf{1.0000} & \textbf{0.9970} & 0.9986 \\

\rowcolor{black!22}
L3$\to$Q1.5
& \textbf{1.0000} & 0.9910 & 0.9989
& \textbf{1.0000} & 0.9910 & 0.9989
& \textbf{1.0000} & 0.9920 & 0.9988
& \textbf{1.0000} & 0.9930 & \textbf{0.9993}
& \textbf{1.0000} & 0.9950 & 0.9990
& \textbf{1.0000} & \textbf{0.9970} & 0.9986 \\

\bottomrule
\end{tabular}%
}
\vspace{-0.08in}
\label{tab:sub_ratio_math}
\end{table*}

\begin{table*}[t]
\centering
\small
\setlength{\tabcolsep}{2.2pt}
\renewcommand{\arraystretch}{1.06}

\caption{Sensitivity to \textbf{Substitution Ratio} on \textbf{General}.
We report TPR@FPR=1\%, ACC@FPR=1\%, and AUC.
Q7/Q3/Q1.5 = Qwen-7B/3B/1.5B, L3 = LLaMA-3B.
Best is bold.}
\label{tab:sub_ratio_general}
\setlength{\tabcolsep}{2.2pt}
\renewcommand{\arraystretch}{1.06}
\resizebox{\linewidth}{!}{%
\begin{tabular}{lccc ccc ccc ccc ccc ccc}
\toprule
& \multicolumn{3}{c}{\textbf{5\%}}
& \multicolumn{3}{c}{\textbf{10\%}}
& \multicolumn{3}{c}{\textbf{15\%}}
& \multicolumn{3}{c}{\textbf{25\%}}
& \multicolumn{3}{c}{\textbf{50\%}}
& \multicolumn{3}{c}{\textbf{65\%}} \\
\cmidrule(lr){2-4}\cmidrule(lr){5-7}\cmidrule(lr){8-10}
\cmidrule(lr){11-13}\cmidrule(lr){14-16}\cmidrule(lr){17-19}
\textbf{Setting}
& \textbf{TPR} & \textbf{ACC} & \textbf{AUC}
& \textbf{TPR} & \textbf{ACC} & \textbf{AUC}
& \textbf{TPR} & \textbf{ACC} & \textbf{AUC}
& \textbf{TPR} & \textbf{ACC} & \textbf{AUC}
& \textbf{TPR} & \textbf{ACC} & \textbf{AUC}
& \textbf{TPR} & \textbf{ACC} & \textbf{AUC} \\
\midrule

\rowcolor{black!7}
Q7$\to$Q3
& \textbf{1.0000} & 0.9910 & 0.9999
& \textbf{1.0000} & 0.9910 & 0.9999
& \textbf{1.0000} & 0.9920 & 0.9999
& \textbf{1.0000} & 0.9930 & 0.9999
& \textbf{1.0000} & 0.9950 & \textbf{1.0000}
& \textbf{1.0000} & \textbf{0.9970} & 0.9999 \\

\rowcolor{black!8}
Q7$\to$Q1.5
& 0.8200 & 0.9820 & 0.9958
& 0.8400 & 0.9750 & 0.9931
& 0.8800 & 0.9740 & 0.9951
& 0.9840 & \textbf{0.9890} & 0.9965
& 0.9860 & 0.9880 & \textbf{0.9971}
& \textbf{0.9877} & \textbf{0.9890} & 0.9969 \\

\rowcolor{black!8}
Q3$\to$Q1.5
& \textbf{1.0000} & 0.9910 & \textbf{1.0000}
& \textbf{1.0000} & 0.9910 & \textbf{1.0000}
& \textbf{1.0000} & 0.9920 & \textbf{1.0000}
& \textbf{1.0000} & 0.9930 & \textbf{1.0000}
& 0.9980 & 0.9940 & \textbf{1.0000}
& 0.9985 & \textbf{0.9960} & \textbf{1.0000} \\

\rowcolor{black!16}
Q3$\to$L3
& \textbf{1.0000} & 0.9910 & \textbf{1.0000}
& \textbf{1.0000} & 0.9910 & 0.9999
& \textbf{1.0000} & 0.9920 & 0.9999
& \textbf{1.0000} & 0.9930 & 0.9999
& \textbf{1.0000} & 0.9950 & 0.9999
& \textbf{1.0000} & \textbf{0.9970} & 0.9999 \\

\rowcolor{black!16}
L3$\to$Q3
& \textbf{1.0000} & 0.9910 & \textbf{1.0000}
& \textbf{1.0000} & 0.9910 & \textbf{1.0000}
& \textbf{1.0000} & 0.9920 & \textbf{1.0000}
& \textbf{1.0000} & 0.9930 & \textbf{1.0000}
& \textbf{1.0000} & 0.9950 & \textbf{1.0000}
& \textbf{1.0000} & \textbf{0.9970} & \textbf{1.0000} \\

\rowcolor{black!22}
Q7$\to$L3
& \textbf{1.0000} & 0.9910 & \textbf{0.9996}
& \textbf{1.0000} & 0.9910 & \textbf{0.9996}
& \textbf{1.0000} & \textbf{0.9920} & \textbf{0.9996}
& 0.9880 & 0.9900 & 0.9995
& 0.9840 & 0.9870 & 0.9990
& 0.9862 & 0.9880 & 0.9986 \\

\rowcolor{black!22}
L3$\to$Q1.5
& \textbf{1.0000} & \textbf{0.9910} & \textbf{1.0000}
& 0.9700 & 0.9880 & 0.9990
& 0.9733 & 0.9880 & 0.9990
& 0.9640 & 0.9840 & 0.9987
& 0.9580 & 0.9740 & 0.9990
& 0.9600 & 0.9710 & 0.9990 \\

\bottomrule
\end{tabular}%
}
\end{table*}

\begin{figure*}[ht]
    \centering
    \includegraphics[width=\linewidth]{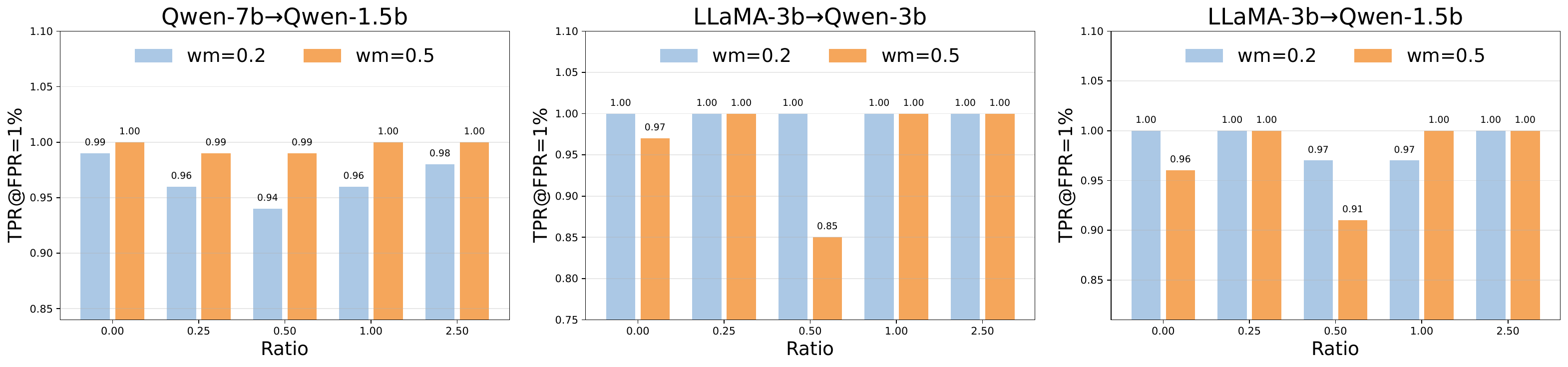}
    
    \caption{Effect of training objective components $\lambda_{\mathrm{sep}}$ and $\lambda_{\mathrm{wm}}$ on substitution performance on \textbf{Code} dataset. Ratio is defined as $\lambda_{\mathrm{sep}}/\lambda_{\mathrm{wm}}$. We report TPR@FPR = 1\% as the score.}
    \label{fig:para_loss_1}
  
\end{figure*}

\textbf{Loss design.} We study the sensitivity of our method to the weights of the watermark loss $\lambda_{\mathrm{wm}}$ and the separability loss $\lambda_{\mathrm{sep}}$ across three types of substitution. We report TPR@FPR = 1\%, with the ratio defined as $\lambda_{\mathrm{sep}}/\lambda_{\mathrm{wm}}$.
Figure~\ref{fig:para_loss_1} shows TPR@1\%
for one Code example from each substitution type, and Figure 6 reports AUC
for the same pairs. Figures~\ref{fig:para_loss_general_auc} and ~\ref{fig:para_loss_general_tpr} report paired AUC and TPR@1\% results
for four representative General pairs, covering changes in size, family,
and both, together with an additional parameter-sensitive case. For completeness, we report the remaining TPR@1\% results on the Code dataset in Figure~\ref{fig:para_loss_code_tpr}.

First, comparing watermark strengths of $0.2$ and $0.5$, we observe that a
larger $\lambda_{\mathrm{wm}}$ can improve detection, although the benefit
depends on the model pair and loss ratio. Second, many substitution settings
achieve high performance at larger loss ratios. These trends support the
usefulness of balancing the watermark and separability terms, but the response
is not uniformly monotonic. In particular, General Qwen-7B$\rightarrow$Qwen-1.5B
remains sensitive, with TPR@1\% of $0.57$ at
$\lambda_{\mathrm{wm}}=\lambda_{\mathrm{sep}}=0.5$.

\begin{figure*}[ht]
    \centering
    \includegraphics[width=\linewidth]{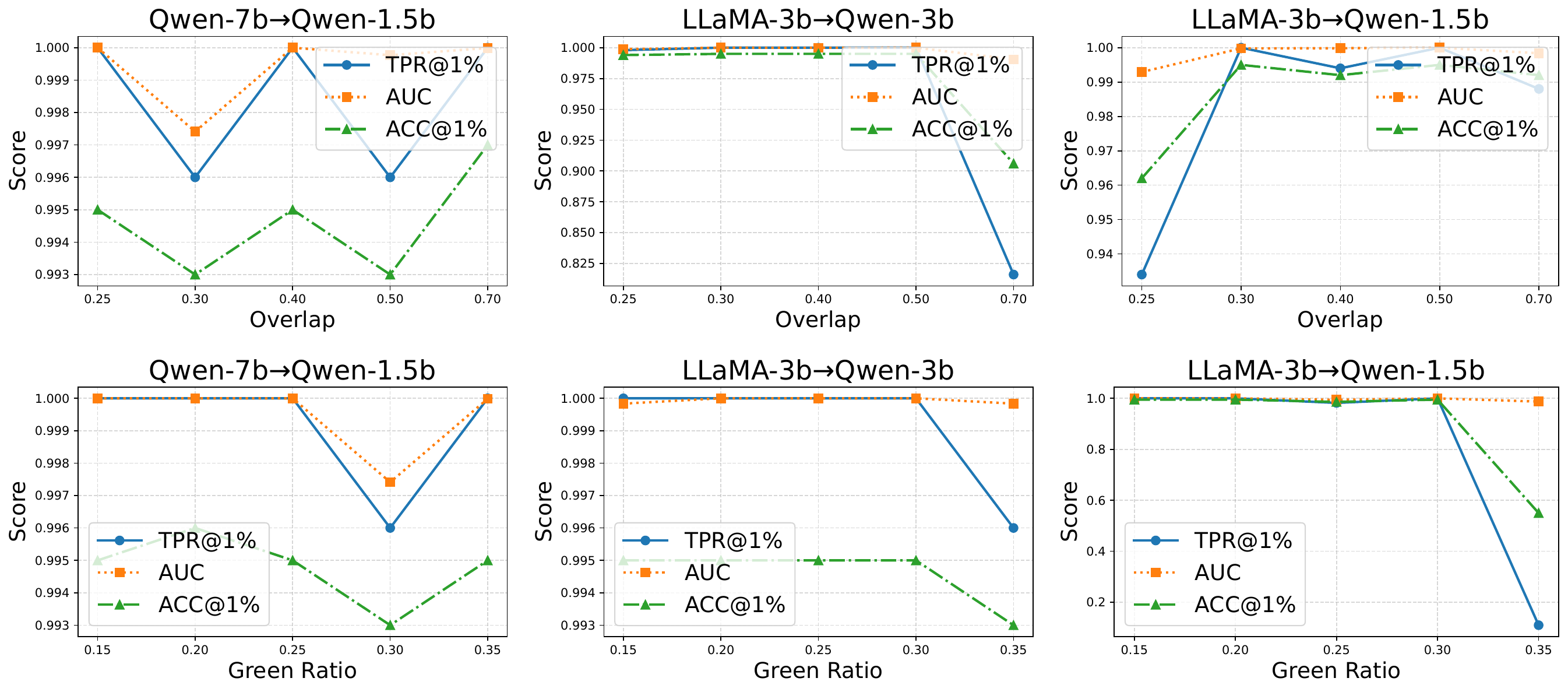}
    \caption{Effect of green ratio $\rho$ and overlap $\tau$ on substitution performance on \textbf{Math} dataset.}
    \label{fig:para_wm}
\end{figure*}

\textbf{Watermark design.}
We study the sensitivity of substitution detection performance to the green ratio $\rho$ and the overlap $\tau$. In Figure~\ref{fig:para_wm}, the first row fixes $\rho$ = 0.3 and varies $\tau$, while the second row fixes $\tau$= 0.3 and varies $\rho$.
Most tested configurations maintain high detection, although
particular model pairs are sensitive: LLaMA-3B$\rightarrow$Qwen-1.5B reaches
TPR@1\% of $0.11$ at $\rho=0.35$, and LLaMA-3B$\rightarrow$Qwen-3B
reaches $0.816$ at $\tau=0.7$ in Figure~\ref{fig:overlap_math} and Figure~\ref{fig:green_math}. More importantly, as discussed in Equation~\ref{eq:capacity}, the feasible
combinations of $\rho$ and $\tau$ are constrained by model identity capacity.
Choosing overly small overlap limits the number of supported identities,
while a small green ratio may restrict the effective vocabulary during summary
generation, creating a potential trade-off with summary quality. We therefore
use $\rho=\tau=0.3$ as a practical default motivated by these design considerations.

\begin{figure*}[ht]
    \centering
    \includegraphics[width=\linewidth]{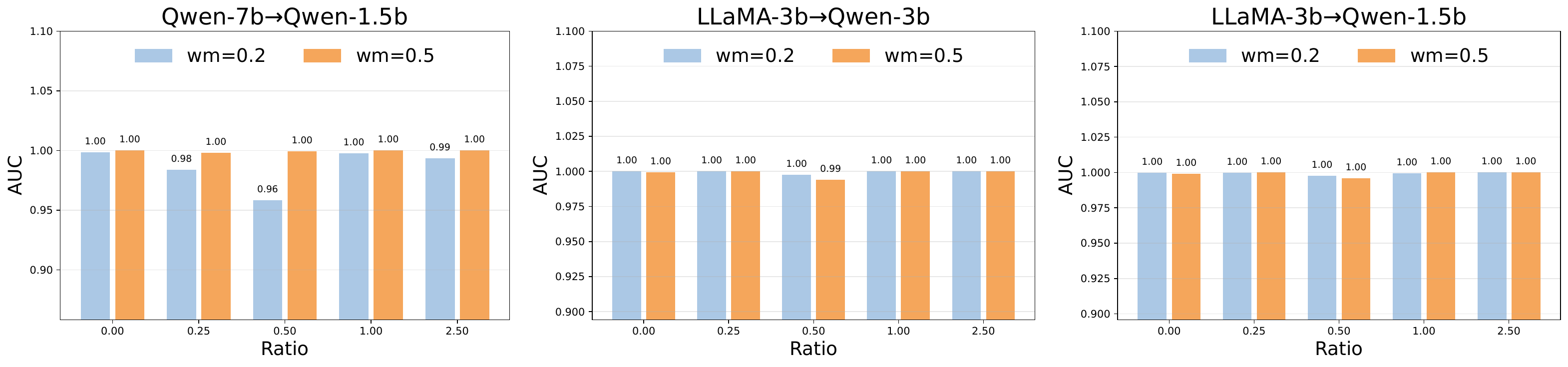}
    \caption{Effect of training objective components $\lambda_{\mathrm{sep}}$ and $\lambda_{\mathrm{wm}}$ on substitution performance on \textbf{Code} dataset. Ratio is defined as $\lambda_{\mathrm{sep}}/\lambda_{\mathrm{wm}}$. We report AUC as the score.}
    \label{fig:para_loss_code_auc}
\end{figure*}

\begin{figure*}[ht]
    \centering
    \includegraphics[width=\linewidth]{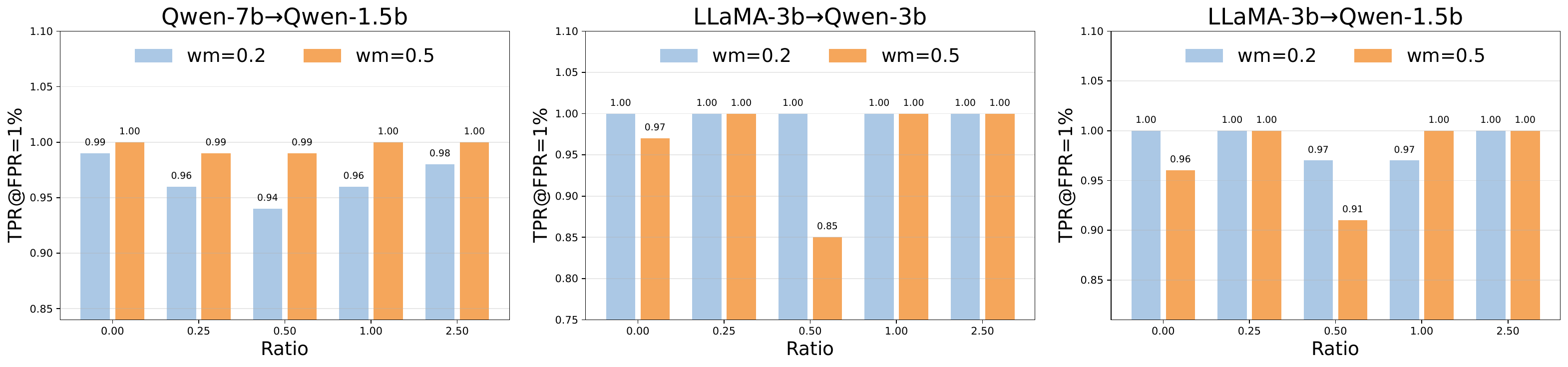}
    \caption{Effect of training objective components $\lambda_{\mathrm{sep}}$ and $\lambda_{\mathrm{wm}}$ on substitution performance on \textbf{Code} dataset. Ratio is defined as $\lambda_{\mathrm{sep}}/\lambda_{\mathrm{wm}}$. We report TPR@FPR = 1\% as the score.}
    \label{fig:para_loss_code_tpr}
\end{figure*}

\begin{figure*}[ht]
    \centering
    \includegraphics[width=0.8\linewidth]{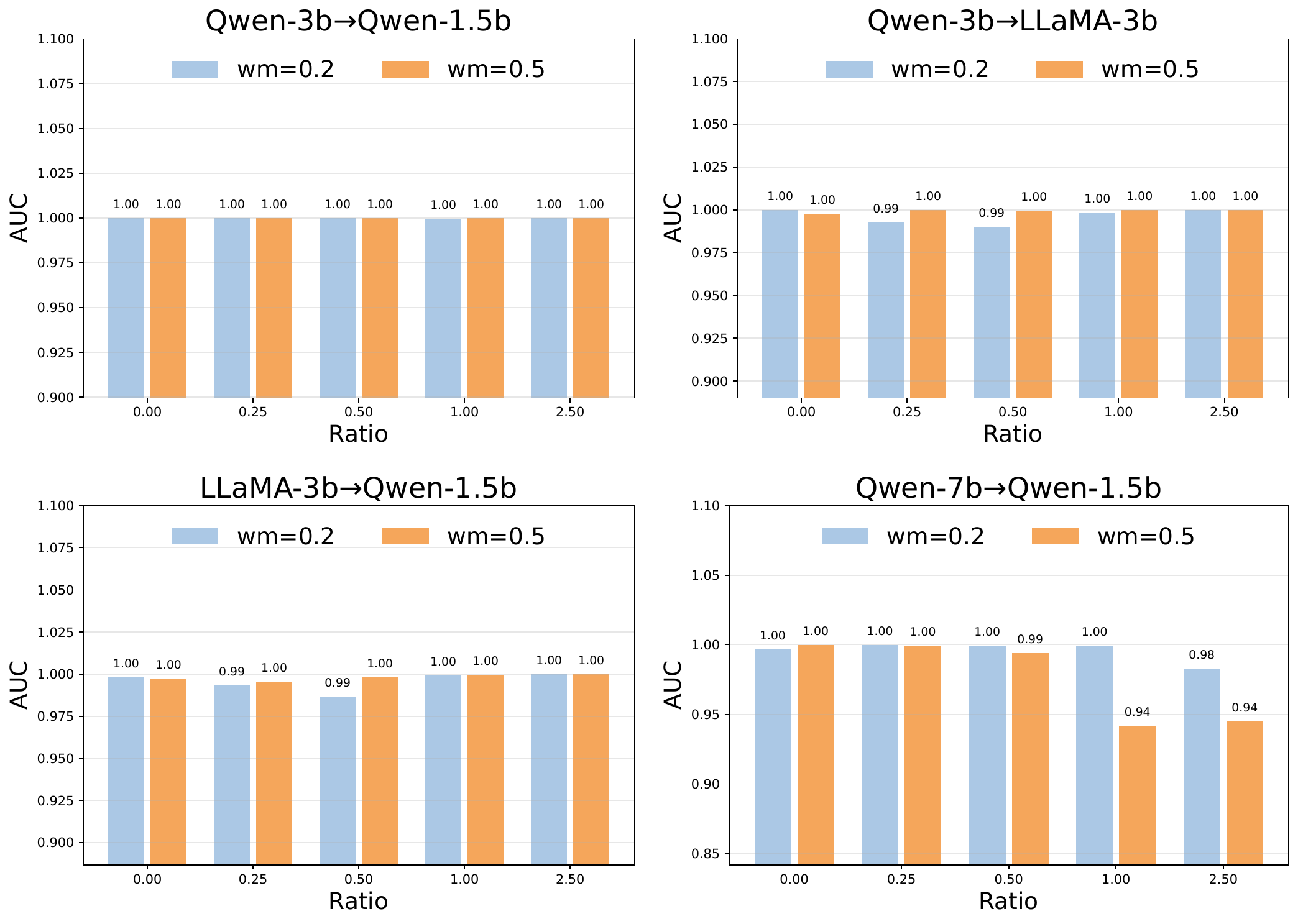}
    \caption{Effect of training objective components $\lambda_{\mathrm{sep}}$ and $\lambda_{\mathrm{wm}}$ on substitution performance on \textbf{General} dataset. Ratio is defined as $\lambda_{\mathrm{sep}}/\lambda_{\mathrm{wm}}$. We report AUC the score.}
    \label{fig:para_loss_general_auc}
\end{figure*}

\begin{figure*}[ht]
    \centering
    \includegraphics[width=0.8\linewidth]{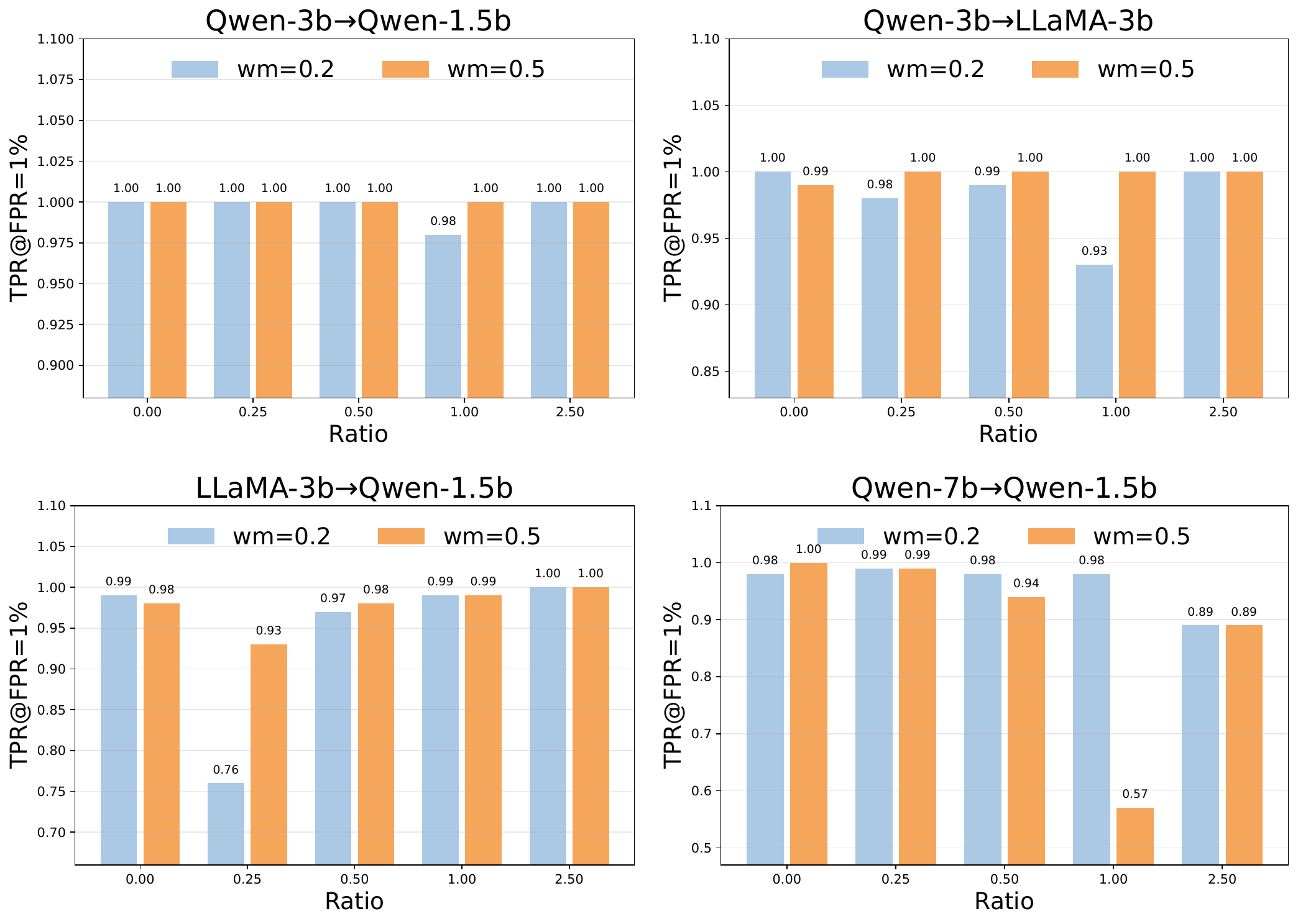}
    \caption{Effect of training objective components on substitution performance on \textbf{General} dataset. We report the TPR@FPR = 1\% as the score.}
    \label{fig:para_loss_general_tpr}
\end{figure*}

\begin{figure*}[th]
    \centering
    \includegraphics[width=\linewidth]{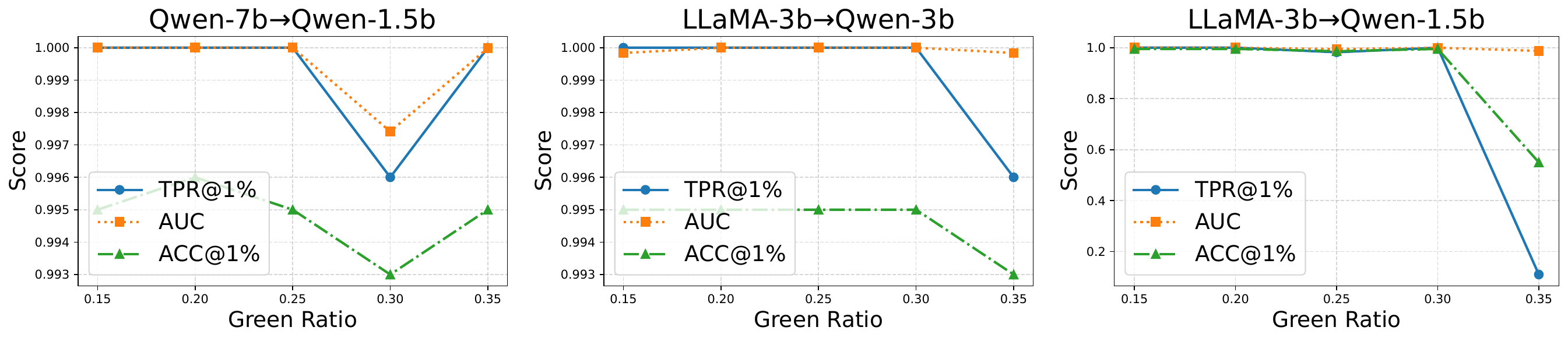}
    \caption{Effect of green ratio $\rho$ on substitution performance on \textbf{Math} dataset.}
    \label{fig:green_math}
\end{figure*}

\begin{figure*}[th]
    \centering
    \includegraphics[width=\linewidth]{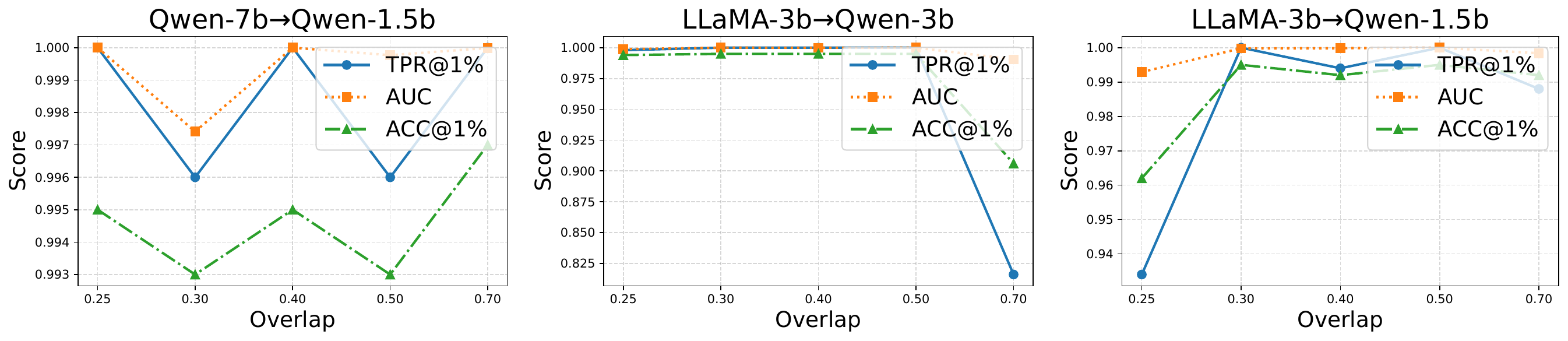}
    \caption{Effect of overlap $\tau$ on substitution performance on \textbf{Math} dataset.}
    \label{fig:overlap_math}
\end{figure*}